\documentclass[review]{elsarticle}
\usepackage{tikz}
\usepackage{pgfplots}
\usepackage{tabu}
\usepackage{ulem}
\usepackage{algorithmic}
\usepackage{graphicx}
\usepackage{textcomp}
\usepackage{xcolor}
\usepackage{subcaption}
\usepackage{multirow}
\usepackage{lipsum}
\usepackage{amsmath,amsfonts,amssymb,latexsym,epsfig,mathtools,appendix}
\usepackage{algorithm}
\def\BibTeX{{\rm B\kern-.05em{\sc i\kern-.025em b}\kern-.08em
    T\kern-.1667em\lower.7ex\hbox{E}\kern-.125emX}}
\usepackage{afterpage}

\newcommand\btheta{\boldsymbol{\theta}}
\newcommand\bx{\boldsymbol{x}}

\newcommand\by{\boldsymbol{y}}

\newcommand\bv{\boldsymbol{v}}
\newcommand\br{\boldsymbol{r}}
\newcommand\bC{\boldsymbol{C}}
\newcommand\bJ{\boldsymbol{J}}
\newcommand\bR{\boldsymbol{R}}
\newcommand\bV{\boldsymbol{V}}
\newcommand\bK{\boldsymbol{K}}

\newcommand\real{\mathbb{R}}
\newcommand\naturals{\mathbb{N}}

\newtheorem{Assumption}{Assumption}

\newtheorem{Remark}{Remark}
\newtheorem{Proposition}{Proposition}
\newcommand{\QED}{\Box}

\usepackage[acronym]{glossaries}
\newacronym{smc}{SMC}{sequential Monte Carlo}
\newacronym{ekf}{EKF}{extended Kalman filter}
\newacronym{enkf}{EnKF}{ensemble Kalman filter}
\newacronym{ukf}{UKF}{unscented Kalman filter}
\newacronym{ckf}{CKF}{cubature Kalman filter}
\newacronym{nhf}{NHF}{nested hybrid filter}
\newacronym{npf}{NPF}{nested particle filter}
\newacronym{smc2}{SMC$^2$}{sequential Monte Carlo square}
\newacronym{pmcmc}{PMCMC}{particle Markov chain Monte Carlo}
\newacronym{pf}{PF}{particle filter}
\newacronym{pdf}{pdf}{probability density function}
\newacronym{pl}{PL}{particle learning}
\newacronym{wrt}{w.r.t.}{with respect to}
\begin{document}

\begin{frontmatter}
	
\title{Nested Gaussian filters for recursive Bayesian inference and nonlinear tracking in state space models}
%\tnotetext[t1]{Agencia Estatal de Investigaci\'on} of Spain (RTI2018-099655-B-I00 CLARA), the regional government of Madrid (project no. Y2018/TCS-4705 PRACTICO) and the Office of Naval Research (award no. N00014-19-1-2226).}

\author{Sara P\'erez-Vieites}
\ead{saperezv@pa.uc3m.es}

\author{Joaqu\'in M\'iguez}
\ead{joaquin.miguez@uc3m.es}

\address{Department of Signal Theory \& Communications, Universidad Carlos III de Madrid. Avenida de la Universidad 30, 28911 Legan\'es, Madrid, Spain.}

%\address{Department of Signal Theory \& Communications, Universidad Carlos III de Madrid. Avenida de la Universidad 30, 28911 Legan\'es, Madrid, Spain.}
%\maketitle

\begin{abstract}
We introduce a new sequential methodology to calibrate the fixed parameters and track the stochastic dynamical variables of a state-space system. The proposed method is based on the nested hybrid filtering (NHF) framework of \cite{Perez-Vieites18}, that combines two layers of filters, one inside the other, to compute the joint posterior probability distribution of the static parameters and the state variables. In particular, we explore the use of deterministic sampling techniques for Gaussian approximation in the first layer of the algorithm, instead of the Monte Carlo methods employed in the original procedure. The resulting scheme reduces the computational cost and so makes the algorithms potentially better-suited for high-dimensional state and parameter spaces. We describe a specific instance of the new method and then study its performance and efficiency of the resulting algorithms for a stochastic Lorenz 63 model with uncertain parameters.
%present numerical results for a stochastic Lorenz 63 model.
\end{abstract}


\end{frontmatter}

\section{Introduction}

State-space models are a popular tool in many fields of science and engineering where researchers and practitioners deal with uncertainty in dynamical systems. A typical state space model consists of:
\begin{itemize}
	\item A random sequence of state vectors, $\bx_t$, that contain the variables of interest for the description of the real-world system at hand, but cannot be observed (at least completely).
	\item A random sequence of noisy observation vectors, $\by_t$, where each $\by_t$ can be related to the state $\bx_t$ through some conditional probability distribution.
	\item A vector $\btheta$ of static model parameters that determine the model behaviour and, typically, have to be estimated from the available data.
\end{itemize}
%State-space dynamical models are used to describe many problems of science, such as statistics, geography and physics. The behaviour of these time-varying state variables also depends on some fixed parameters, that need to be estimated. Then, we are interested in tracking both dynamical state variables and fixed parameters by using long sequences of observations. 

%We address the problem of tracking the time evolution of state-space dynamical systems, whose behaviour relies on a set of unknown fixed parameters. Then, both time-varying states and static parameters need to be estimated, and for this purpose long observation sequences are collected. 

%Generally, jointly carrying out Bayesian model calibration (parameter estimation) and filtering or data assimilation (state tracking) poses several practical and theoretical difficulties. 

	Classical filtering methods \cite{Anderson79,Gordon93,Doucet00,Julier00,Djuric03,Ristic04}, including both Kalman-based algorithms and Monte Carlo schemes (particle filters \cite{Djuric03}) tackle the problem of predicting and tracking the states $\bx_t$ using the observations $\by_t$, while assuming that the parameters $\btheta$ are given. This is hardly ever the case in practice, though, and the fixed parameters $\btheta$ have to be estimated from the data $\by_t$ as well. The joint tracking of $\bx_t$ and estimation of $\btheta$ involves several practical and theoretical difficulties.
	The straightforward approach is the use of state augmentation \cite{Lourens12,Hassanzadeh08,Liu01,Andrieu04,Kitagawa98,Zhang17}, where an extended state is introduced that includes both the static parameters $\btheta$ and the dynamical variables $\bx_t$. This methodology can be applied with any standard filtering technique such as Kalman-like methods (either extended \cite{Lourens12} or sigma-point-based \cite{Hassanzadeh08} approximations) and \glspl{pf} \cite{Liu01,Kitagawa98,Liu01b}. In the case of \glspl{pf}, artificial dynamics are usually introduced for the fixed parameters, reinterpreting them as slow-changing dynamical variables in order to avoid the degeneracy of the Monte Carlo approximation. Both with Kalman and particle filtering techniques, state augmentation is easy to apply but the resulting methods are often inefficient and lack theoretical guarantees.
%	In general, the parameter estimation and the state tracking pose several difficulties when they are carried out jointly, both practical and theoretical. 
%Over the years, many procedures have been proposed. On the one hand, we can use state augmentation methods \cite{Liu01,Andrieu04,Kitagawa98,Zhang17}, where we introduce an extended state that includes both static parameters and dynamical variables. This methodology can be applied with any standard filtering technique. However, the parameters need to be reinterpreted as slow changing state variables, introducing artificial dynamics in order to explore appropriately the parameter space. This introduces a bias in the results that is difficult to quantify. 
When the posterior distribution of the static parameters can be represented by a set of finite statistics, classical state-augmentation can be replaced by a two-stage procedure where one first samples the posterior of the parameters and then the states (conditional on the parameters). Such methods are often referred to as particle learning \cite{Carvalho10,Storvik02,Djuric02,Nemeth13}. The assumption of having a description of the posterior distribution of the parameters is rather restrictive, though. A more general strategy is the use of recursive maximum likelihood methods \cite{Kantas15,Andrieu04,Andrieu12,Tadic10,Ding14}. These techniques are well-principled and can be applied to a broad class of models. However, they provide point estimates of the unknowns as the observations are collected and not full posterior distributions. Therefore, the uncertainty is not quantified.

On the other hand, some major advances in the last few years have led to well-principled algorithms that can solve (numerically) the joint model inference (parameter estimation) and state tracking problems. They are fundamentally Bayesian methods that aim at computing the posterior probability distribution of the unknown states and parameters given sequentially collected observations. This approach not only provides point estimates but also information about the uncertainty of those estimates. Some examples are the \gls{smc2} \cite{Chopin12}, the \gls{pmcmc} \cite{Andrieu10} and the \gls{npf} \cite{Crisan18bernoulli} methods. However, both \gls{smc2}  and \gls{pmcmc} are batch (non recursive) techniques. Therefore, every time a new observation is introduced, the whole sequence of observations may need to be processed in order to compute the new Bayes estimator.
%major breakthroughs have been attained in the last few years, including methods such as sequential Monte Carlo square (SMC$^2$) \cite{Chopin12} or particle Markov chain Monte Carlo (PMCMC) \cite{Andrieu10}. They aim at computing the posterior probability distribution of all the unknown variables and parameters of the system \textcolor{blue}{given data that is sequentially collected.} However, these are batch techniques, i.e., the whole sequence of observations may have to be processed every time a new observation arrives in order to obtain estimates. 
A closely-related methodology that is better suited for long sequences of observed data is the \gls{npf} \cite{Crisan18bernoulli,Crisan17}. It applies the same principles as \gls{smc2}, but \glspl{npf}  are purely recursive. It is a scheme with two intertwined layers of Monte Carlo methods, one inside the other, using the ``inner'' layer to track the dynamic state variables and the ``outer'' layer for parameter estimation. However, since the \gls{npf} uses Monte Carlo in both layers of filters, its computational cost becomes prohibitive in high-dimensional problems. With the aim of reducing this cost, the class of \glspl{nhf} was introduced in \cite{Perez-Vieites18}. \glspl{nhf} are recursive algorithms with the same multi-layer structure as \glspl{npf} but they enable the use of non-Monte Carlo filtering techniques in the ``inner'' layer (state tracking). Therefore, an \gls{nhf}  is a general scheme that can approximate the posterior probability distribution of the static parameters with a recursive Monte Carlo or quasi-Monte Carlo \cite{Gerber15} method and combine it with different filtering (Monte Carlo or Kalman-based) techniques in order to approximate the posterior probability distribution of the dynamical state variables of the system.

%For these settings, nested hybrid filters (NHF's) \cite{Perez-Vieites18} are more appealing, since they introduce Gaussian filtering techniques in the second layer of the algorithm, which reduces the computational cost. 

%This methodology is better suited for long sequences of observations, although its computational cost is still prohibitive in high-dimensional problems as it uses Monte Carlo in both layers of filters. 

% In this paper, we introduce a further generalization of the NHF methodology by describing how to apply non-Monte Carlo methods in the first layer of the algorithm. This allows the combination of virtually any type of Gaussian or particle filter in any of the two layers of the nested structure. In this Bayesian approach, we approximate the posterior probability distribution of the unknowns (the parameters and the state variables) given the sequential available data. To be specific, we show in detail how to obtain a NHF that employs a deterministic-sampling Gaussian approximation (such as the cubature Kalman filter (CKF) \cite{Arasaratnam09} or the unscented Kalman filter (UKF) \cite{Julier00}) in the first (parameter) layer with an extended Kalman filter (EKF) in the second (state) layer. 
In this paper, we extend the \gls{nhf} methodology to enable the use of non-Monte Carlo schemes in both layers of the nested filtering procedure. The new scheme, therefore, is a methodological generalization of the algorithms in \cite{Perez-Vieites18,Crisan18bernoulli,Crisan17} that comprises a broad class of nested filters for which it is possible to use and combine Gaussian or particle filters in any of the two layers. The new algorithms remain purely recursive and yield numerical approximations of the posterior probability of the unknown state variables and parameters using the sequentially collected observations.
 	%introduce a new methodology that describes how to apply non-Monte Carlo methods in both the first and the second layer of the algorithm. This is a further generalization of the \gls{nhf}. Within this nested structure, the use and combination of any type of Gaussian or particle filter in any of the two layers is allowed. It is a Bayesian approach, where the posterior probability of the unknown state variables and parameters is approximated by using the sequential available observations. 

To be specific, in this work we explain in detail the use of a deterministic-sampling Gaussian approximation (such as the \gls{ukf} \cite{Julier00} or the \gls{ckf} \cite{Arasaratnam09}) in the outer layer of the nested filtering scheme. Either particle of Gaussian (Kalman-based) filters can be easily plugged into the inner layer (we implement extended Kalman filters in our experiments for simplicity). The key difficulty to be tackled when using non-Monte Carlo methods in the outer layer is to keep the algorithm recursive. This was achieved for the Monte Carlo methods in \cite{Crisan18bernoulli} and \cite{Perez-Vieites18} using a ``jittering'' procedure that cannot be extended to Gaussian filters in a practical way. Instead, we place a condition on the update of the filter in the outer layer that depends on a distance defined on the parameter space. When the distance between consecutive parameter updates falls below a prescribed threshold the algorithm operates in a purely recursive manner. This approach can work adequately when the posterior probability distributions of the state variables are continuous \gls{wrt} the unknown parameters, and we prove that this is the case under regularity assumptions on the state-space model.
	%In addition, for the state tracking, we use an \gls{ekf} in the second layer. Although \gls{nhf}  works recursively when Monte Carlo methods are used in the first layer, this is not straightforward when we replace them with Gaussian techniques. For this reason, we evaluate if the algorithm can work recursively in every time step by using a deciding threshold. This implies the approximation of the posterior probability distribution of the state variables given the parameters, that is only computed when the error introduced is minimal. Then, the algorithm not only remains recursive as \gls{nhf} , but we also reduce the computational cost since we need considerably less sigma-points (\gls{ukf}) or cubature points (\gls{ckf}) than particles when we use Monte Carlo methods. 

	In order to assess the performance of the proposed nested methods we have implemented a recursive scheme that employs a \gls{ukf} in the outer layer (for parameter estimation) and a bank of \glspl{ekf} in the inner layer (for state tracking). We have carried out a simulation study to compare the performance of this algorithm with two state-augmented Gaussian filters (a \gls{ukf} and a \gls{enkf} \cite{Evensen03}) as well as another nested algorithm that combines a particle filter in the outer layer with \glspl{ekf} in the inner layer \cite{Perez-Vieites18}. The methods are applied to the problem of tracking a stochastic Lorenz 63 model with three unknown parameters in the state equation.

	The rest of the paper is organized as follows. In Section \ref{sProblemStatement} we describe the class of state-space models with unknown parameters to be studied through the paper. In Section \ref{sNestedGaussianFilters} we derive the family of nested Gaussian filters with sigma-point approximations in the outer layer. Our computer simulation results are presented in Section \ref{sExample} and, finally, Section \ref{sConclusions} is devoted to the conclusions.
%We state the problem to be addressed in Section \ref{sProblemStatement}. In Section \ref{sNestedGaussianFilters}, we describe the new methodology and how it can be made to work recursively. Some numerical results for the stochastic Lorenz 63 model are shown in Section \ref{sExample} and conclusions are drawn in Section \ref{sConclusions}.

\section{Problem Statement} \label{sProblemStatement}

\subsection{State space models} \label{ssStatespacemodels}

We are interested in the class of Markov state-space dynamical systems with additive noise that can be described by the pair of equations 
\begin{align}
	\bx_t &= f(\bx_{t-1}, \btheta) + \bv_t, \label{eqxdiscrete}\\
	\by_t &= g(\bx_t,\btheta) + \br_t, \label{eqydiscrete}
\end{align}
 where $t \in \naturals$ denotes discrete time, $\bx_t \in \mathbb{R}^{d_x}$ is the $d_x$-dimensional system state, $f \colon \mathbb{R}^{d_x} \times \mathbb{R}^{d_{\theta}} \longrightarrow \mathbb{R}^{d_x}$ and $g \colon \mathbb{R}^{d_x} \times \mathbb{R}^{d_{\theta}} \longrightarrow \mathbb{R}^{d_y}$, $d_x \ge d_y$, are possibly nonlinear functions parameterized by a (random but fixed) vector of unknown parameters, $\btheta \in \mathbb{R}^{d_{\theta}}$, $\by_t \in \mathbb{R}^{d_y}$ is the observation vector at time $t$ and $\bv_t$ and $\br_t$ are zero-mean random vectors playing the roles of state and observations noises.
%  with zero mean and covariance matrices $\bV$ and $\bR$, respectively.

The system of equations \eqref{eqxdiscrete} and \eqref{eqydiscrete} can be described in terms of a set of relevant \glspl{pdf}\footnote{We adopt an argument-wise notation for \glspl{pdf}. If we have two random variables $\bx$ and $\by$, we write $p(\bx)$ and $p(\by)$ for their respective \glspl{pdf}, which are possibly different. In a similar way, $p(\bx,\by)$ denotes the joint \gls{pdf} of the two random variables and $p(\bx|\by)$ denotes the conditional \gls{pdf} of $\bx$ given $\by$.}, specifically
\begin{align}
\bx_0 &\sim p(\bx_0), \quad \btheta \sim p(\btheta), \label{eqX0}\\ 
 %\label{eqTheta} \\
&\bx_t \sim p(\bx_t | \bx_{t-1}, \btheta), \label{eqxt}\\
&\by_t \sim p(\by_t | \bx_t, \btheta), \label{eqyt}
\end{align}
where $p(\btheta)$ and $p(\bx_0)$ are the a \textit{priori} \glspl{pdf} of the parameters and the state, respectively, $p(\bx_t | \bx_{t-1}, \btheta)$ is the conditional density of the state $\bx_t$ given $\bx_{t-1}$ and the parameter vector $\btheta$, and $p(\by_t | \bx_t, \btheta)$ is the conditional \gls{pdf} of the observation $\by_t$ given $\bx_t$ and $\btheta$. We assume that $\by_t$ is conditionally independent of all other observations (given $\bx_t$ and $\btheta$) and the prior \glspl{pdf} of the state, $p(\bx_0)$, and the parameters, $p(\btheta)$, are known and the corresponding probability distributions are independent.

%\begin{itemize}
%	%\item $t \in \naturals$ denotes discrete time;
%	%\item $\bx_t$ is the $d_x$-dimensional (random) state vector at time $t$, taking values in the state space $\mathcal{X} \subseteq \real^{d_x}$; 
%	%\item $\btheta$ is the $d_{\theta}$-dimensional vector of fixed parameters;
%	\item $p(\btheta)$ and $p(\bx_0)$ are the a \textit{priori} pdfs of the parameters and the state;
%	\item $p(\bx_t | \bx_{t-1}, \btheta)$ is the conditional density of the state $\bx_t$ given $\bx_{t-1} = \bx_{t-1}$ and the parameter vector $\btheta = \btheta$;
%	\item $\by_t$ is the $d_y$-dimensional observation vector at time $t$, taking values in the observation space $\mathcal{Y} \subseteq \real^{d_y}$. We assume $\by_t$ is conditionally independent of all other observations given $\bX_t$ and $\Theta$;
%	\item $p(\by_t | \bx_t, \btheta)$ is the conditional pdf of $\by_t$ given $\bx_t = \bx_t$ and $\btheta = \btheta$.
%\end{itemize}
%%A broad class of systems can be described by the model in \eqref{eqX0}--\eqref{eqyt}, both linear and nonlinear, with Gaussian or non-Gaussian perturbations. 
%We assume that the prior distributions of the state, $p(\bx_0)$, and the parameters, $p(\btheta)$, are known, and we aim at estimating both $\Theta$ and $\bX_t$ recursively.

\subsection{Model inference} \label{ssModelinference}
The key difficulty in this class of models is the Bayesian estimation of the parameter vector $\btheta$, since its calibration is necessary in order to track the state variables and predict the evolution of the system. From the viewpoint of Bayesian analysis, we aim at computing the posterior pdf $p(\btheta | \by_{1:t})$ as it contains all the relevant information for the estimation task at discrete time $t$. However, this pdf can be written as
\begin{equation}
p(\btheta | \by_{1:t}) = \int p(\btheta, \bx_t | \by_{1:t}) d\bx_t,
\end{equation} 
leading naturally to approximations for $p(\btheta, \bx_t | \by_{1:t})$ for each $t$. This means that when computing $p(\btheta|\by_{1:t})$ we may not only estimate the parameter vector $\btheta$, but we may also implicitly track the state dynamical variables. The main aim of this paper is to obtain a Gaussian approximation of $p(\btheta | \by_{1:t})$ within a nested Gaussian filtering scheme, whose second layer of filters will provide, in addition, Gaussian approximations for $p(\bx_t | \by_{1:t}, \btheta)$.

\section{Nested Gaussian filters} \label{sNestedGaussianFilters}

In this section, we introduce a class of nested filter for state-space models with unknown parameters that combine different types of Gaussian approximations in the inner and outer layers. We outline the methodology used to obtain the Gaussian approximations of $p(\btheta | \by_{1:t})$ (in the outer layer) and $p(\bx_t | \by_{1:t}, \btheta)$ (in the inner layer). 

In the sequel we keep using $p(\cdot)$ to denote the actual \glspl{pdf}. We aim, however, at constructing Gaussian approximations of the posterior \glspl{pdf} induced by the state-space model \eqref{eqX0}-\eqref{eqyt} and the sequence of observations. For this purpose, we introduce notation $\mathcal{N}(\bx | \bar{\bx}, \bC)$ to denote the Gaussian \gls{pdf} with mean $\bar{\bx}$ and covariance matrix $\bC$. We will show how to recursively compute approximations $p(\btheta | \by_{1:t}) \simeq \mathcal{N}(\btheta| \hat{\btheta}_t, \hat{\bC}_t^{\btheta})$, $p(\bx_t | \by_{1:t}, \btheta) \simeq \mathcal{N}(\bx_t |\hat{\bx}_{t,\btheta}, \bC_{t,\btheta}^{\bx})$ and $p(\bx_t | \by_{1:t}) \simeq \mathcal{N}(\bx_t | \hat{\bx}_t, \bC_t^{\bx})$.

%an specific case of a nested Gaussian filter, considering an UKF in the first layer, for the approximation of the posterior distribution of the parameters, and an ensemble of EKFs approximating the posterior distribution of the state. For the UKF, $M=2d_\theta +1$ sigma-points are generated at time $t$, $\btheta^i_t$, $i=\{1,\ldots,M\}$. For each of them, an extended Kalman filter tracks the state variables until next observation and computes $p(\by_t|\by_{1:t-1},\btheta^i_t)$. With this and following the derivations explained below, we can update the parameter estimates and their covariance matrix. From these estimates, new sigma-points are generated and we repeat the procedure. 

\subsection{Sequential Gaussian approximation} \label{ssMethodology}

Let us aim at computing expectations of the form $ \mathbb{E}[f(\btheta)| \by_{1:t}] = \int f(\btheta) p(\btheta | \by_{1:t}) d\btheta$ for some test function of the parameters, $f(\btheta)$. Using Bayes' rule, we have
\begin{equation}
p(\btheta|\by_{1:t}) = \frac{p(\by_t | \by_{1:t-1}, \btheta)}{p(\by_t|\by_{1:t-1})} \times p(\btheta | \by_{1:t-1}),
\end{equation}
hence, we can rewrite the posterior expectation as
\begin{equation}
\mathbb{E}[f(\btheta)| \by_{1:t}] = \int \psi(\btheta) p(\btheta | \by_{1:t-1}) d\btheta, \label{eqDerivation}
\end{equation}
where the function $\psi(\btheta)$ is constructed as
\begin{equation}
\psi(\btheta) \coloneqq \frac{f(\btheta) p(\by_t | \by_{1:t-1}, \btheta)}{p(\by_t | \by_{1:t-1})}. \label{eqpsi}
\end{equation}
If we assume that $p(\btheta | \by_{1:t-1})$ is Gaussian, then we can approximate \eqref{eqDerivation} using cubature rules \cite{Arasaratnam09} or the unscented transform (UT) \cite{Julier00}. Specifically, a Gaussian approximation $\mathcal{N}(\btheta | \hat{\btheta}_{t-1}, \bC_{t-1}^{\btheta}) \simeq p(\btheta | \by_{1:t-1})$ can be represented at time $t$ by a set of reference points and weights, $\{\btheta_t^i, w_t^i\}$, $i = 1,\ldots,M$, which in turn we may use to approximate the integral in \eqref{eqDerivation} as
\begin{equation}
	\int \psi(\btheta) p(\btheta | \by_{1:t-1}) d\btheta \simeq \sum_{i=1}^{M} \psi(\btheta_t^i) w_t^i.
\end{equation} 
On the other hand, the pdf in the denominator of expression \eqref{eqpsi}, $p(\by_t | \by_{1:t-1})$, can be written as
\begin{equation}
p(\by_t | \by_{1:t-1}) = \int p(\by_t, \btheta | \by_{1:t-1}) d\btheta, \label{eqDen}
\end{equation}
where the joint pdf of $\by_t$ and $\btheta$ given all previous observations can be decomposed as 
\begin{equation}
p(\by_t, \btheta | \by_{1:t-1}) = p(\by_t |\btheta, \by_{1:t-1}) p(\btheta | \by_{1:t-1}).
\end{equation}
Then, the integral in \eqref{eqDen} can also be approximated using the same set of reference points and weights as
\begin{equation}
p(\by_t | \by_{1:t-1}) \simeq \sum_{i=1}^{M} p(\by_t | \by_{1:t-1}, \btheta_t^i) w_t^i. \label{eqDenominator}
\end{equation}
Finally, we can approximate the \gls{pdf} $p(\by_t | \by_{1:t-1}, \btheta_t^i)$, $i = 1,\ldots,M$, using a bank of $M$ Gaussian filters placed in the second layer of the nested filter \cite{Perez-Vieites18}. Once these densities are computed, we can approximate $p(\by_t | \by_{1:t-1})$ as in \eqref{eqDenominator}. 

The argument above enables us to approximate any integral $\int f(\btheta) p(\btheta | \by_{1:t}) d\btheta$. In particular, we can compute the mean vector and covariance matrix of $p(\btheta|\by_{1:t}) \simeq \mathcal{N}(\btheta | \hat{\btheta}_t, \bC_t^{\btheta})$ by taking $f(\btheta)=\btheta$ and $f(\btheta)=(\btheta - \hat{\btheta}_t)(\btheta - \hat{\btheta}_t)^\top$, respectively, where 
\begin{equation}
	\hat{\btheta}_t = \int \btheta p(\btheta | \by_{1:t}) d\btheta.
\end{equation}
Specifically, we obtain the formulation for approximating the mean parameter vector, $\hat{\btheta}_t$, and its covariance matrix, $\hat{\bC}^{\btheta}_t$, sequentially as
\begin{align}
\hat{\btheta}_t &\simeq \sum_{i=1}^{M} {{\btheta}}_t^{i} \frac{p(\by_t | \by_{1:t-1}, {\btheta}_t^{i})}{p(\by_t | \by_{1:t-1})} w_t^{i} \quad \text{and} \label{eqParamEst} \\
\hat{\bC}^{\btheta}_t &\simeq \sum_{i=1}^{M} ({\btheta}_t^{i} - \hat{\btheta}_t) ({\btheta}_t^{i} - \hat{\btheta}_t)^{\top} \frac{p(\by_t | \by_{1:t-1}, {\btheta}_t^{i})}{p(\by_t| \by_{1:t-1})} w_t^{i}. \label{eqCovEst}
\end{align}
We outline the procedure for the sequential computation of the Gaussian approximations $\mathcal{N}(\btheta | \hat{\btheta}_t, \bC_t^{\btheta}) \simeq p(\btheta | \by_{1:t})$, $t=1,2,\ldots$, in Algorithm \ref{alNGF}. The calculations done in the second layer of filters are summarized in step \ref{stepGF}. Notice that, at any time $t \ge1$, we update the reference points $\btheta_{t}^i$, $i=1,\ldots,M$, and, therefore, we need to run the $M$ Gaussian filters in the second layer from scratch (i.e., from $n=0$ to $n=t$) in order to (approximately) evaluate the densities $p(\by_t | \by_{1:t-1}, \btheta_t^i)$. Thus, Algorithm \ref{alNGF} is sequential but not recursive and, as a consequence, not well suited to handle long sequences of observations.

%Explicar de donde sale el denominador
%%%%%%%%%%%%%%%%%%%%%%%%%%%%%%%%%%%%%%%%%%%%%%%%%%%%%%%%%%%
%When $f(\btheta) = \btheta$, we obtain the estimate of the parameters, $\hat{\btheta}_t$, at time $t$. Its approximation is
%\begin{equation}
%\hat{\btheta}_t \approxeq \sum_{i=1}^{t} {{\btheta}}_t^{i} \frac{p(\by_t | \by_{1:t-1}, {\btheta}_t^{i})}{p(\by_t | \by_{1:t-1})} w_t^{i}, \label{eqParamEst} 
%\end{equation}
%where $\btheta^i_t$ and $w^i_t$, $i = \{1,\ldots,M\}$, are respectively the reference points of first layer and their correspondent weights at time $t$. 

%Following the same rule, the approximated estimate of the covariance matrix of the parameters can be expressed as
%\begin{equation}
%\hat{\bC}^\theta_t \approxeq \sum_{i=1}^{t} ({\btheta}_t^{i} - \hat{\btheta}_t) ({\btheta}_t^{i} - \hat{\btheta}_t)^{\top} \frac{p(\by_t | \by_{1:t-1}, {\btheta}_t^{i})}{p(\by_t| \by_{1:t-1})} w_t^{i}. \label{eqCovEst}
%\end{equation}
%\afterpage{
%	\clearpage
%	
\begin{algorithm} \caption{\label{alNGF} Nested Gaussian filters.}
	
	\textbf{Inputs}: 
	\begin{itemize}
		\item[-] Prior \glspl{pdf} $p(\bx_0)$ and $p(\btheta)$. Assume that either $p(\bx_0)$ is Gaussian or a Gaussian approximation is available.
	\end{itemize}
	
	\textbf{Procedure}:
	\quad \begin{enumerate}	
		\item Initialization
		
		\begin{enumerate}
			\item Generate $M$ reference points, $\btheta^i_1$, from $p(\btheta)\simeq \mathcal{N}(\btheta | \btheta_0, \bC_0^{\btheta})$ for $i = 1,\ldots,M$, with weights $w_1^i$.
			%\item Initialize $\bx_0$ from $p(\bx_0)$.
		\end{enumerate}
		
		\item Sequential step, $t \ge 1$.	
		%In this step, it is assumed that the parameter and the state particle sets of the previous step is available for $n \ge 1$. This set have to be updated in the following way:
		\begin{enumerate}
			\item For each $i=1,\ldots,M$, use a Gaussian filter to approximately compute $p(\by_t | \by_{1:t-1}, {\btheta}_t^{i})$. \label{stepGF}
			\item Compute $\hat{\btheta}_t$ and $\hat{\bC}^{\btheta}_t$ via \eqref{eqParamEst} and \eqref{eqCovEst}. \label{stepupdate}
			
			\item Generate new reference points $\btheta^i_{t+1}$ and weights $w_{t+1}^i$, $i = 1,\ldots,M$, from $\hat{\btheta}_t$ and $\hat{\bC}^{\btheta}_{t}$. \label{stepresampling}
			
		\end{enumerate}
	\end{enumerate}
	
	\textbf{Outputs}: $\hat{\btheta}_t$ and $\hat{\bC}_t^{\btheta}$.
\end{algorithm}
%\thispagestyle{empty}
%\clearpage 
%} 
\subsection{Recursive algorithm} \label{ssRecursivealgorithm}

For every new observation vector $\by_t$, the pdf's $p(\by_t | \by_{1:t-1}, \btheta_t^i)$ are computed by running the nested filters from time $0$ until the current time $t$, which makes the computational cost increase with $t^2$. 

However, the entries of the covariance matrix, $\hat{\bC}_t^{\btheta}$, also tend to stabilize over time, which makes the difference between consecutive reference points, $\btheta_{t}^i - \btheta_{t-1}^i$, decrease. If we also assume that the function $p(\by_t | \by_{1:t-1}, \btheta)$ is continuous in $\btheta$, then we can make the computation recursive by assuming that $p(\by_t | \by_{1:t-1}, \btheta_t^i) \simeq p(\by_t | \by_{1:t-1}, \btheta_{t-1}^i )$ when $\btheta_{t}^i \simeq \btheta_{t-1}^i$. For the sake of clarity we summarize the steps for computing $p(\by_t | \by_{1:t-1}, \btheta_t^i)$ in Algorithm \ref{alEKF2ndlayer}, relying on a bank of \glspl{ekf}. Let us remark that the second layer of the nested algorithm can be implemented using a variety of filters, e.g., particle filters as in \cite{Crisan18bernoulli} or Gaussian filters as in \cite{Perez-Vieites18}, including \glspl{ukf} as we have done for the first layer. We choose a bank of \glspl{ekf} simply because it is the computationally less demanding alternative.

Algorithm \ref{alNGFRecursive} outlines a recursive nested Gaussian filter with a \gls{ukf}/\gls{ckf} in the first layer and \glspl{ekf} in the second layer. It can be seen as a recursive and explicit implementation of Algorithm \ref{alNGF}. The initialization remains the same (step \ref{stepInitRefPoint}), computing $M$ reference points $\btheta_1^i$ and weights $w_1^i$, $i=1,\ldots,M$, from the prior $p(\btheta) \simeq \mathcal{N}(\btheta | \btheta_0, \bC_0^{\btheta})$. Also, we initialize the state and its covariance matrix in every Gaussian filter of the second layer (step \ref{stepInitX}) by setting $\hat{\bx}_0^i = \hat{\bx}_0$ and $\hat{\bC}_0^{\bx,i} = \bC_0^{\bx}$, $i = 1,\ldots,M$, from the prior $p(\bx_0) = \mathcal{N} (\bx_0 | \hat{\bx}_0, \bC_0^{\bx})$. 

%To be specific, let $d(\btheta , \btheta')$ denote a distance function in the space of the parameters \footnote{This scheme is explained for a generic distance $d(\btheta_{t}^i, \btheta_{t-1}^i)$ since it works for different types of distances such as
%	\begin{align}
%	d_{\text{euclidean}}(\btheta_{t}^i , \btheta_{t-1}^i) &=\sqrt{ \sum_{j=1}^{d_\theta}(\btheta_{t,j}^i - \btheta_{t-1,j}^i)^2}, \label{eqeuclideandist}\\
%	d_{\text{Chebyshev}}(\btheta_{t}^i , \btheta_{t-1}^i) &= \max_{1\leq j\leq d_{\theta}}|\btheta_{t,j}^i - \btheta_{t-1,j}^i| \quad \text{and} \label{eqChebyshevdist}\\
%	d_{\text{taxicab}}(\btheta_{t}^i , \btheta_{t-1}^i) &= \sum_{j=1}^{d_{\theta}}|\btheta_{t,j}^i - \btheta_{t-1,j}^i|;
%	\end{align}
%	that are the Euclidean, the Chebyshev and the taxicab distances respectively.
%}. 
%We compute $d(\btheta_{t}^i , \btheta_{t-1}^i)$ }and compare it against a prescribed threshold $\lambda > 0$

The sequential procedure starts by approximating $p(\bx_t | \by_{1:t-1}, \btheta_t^i)$ with the second layer of Gaussian filters (step \ref{stepThreshold}). This is done differently depending on whether we assume $\btheta_{t}^i \simeq \btheta_{t-1}^i$ or not. To be specific, the norm\footnote{Although other metrics $d(\btheta_t^i, \btheta_{t-1}^i)$ could be used, we adopt p-norms of the difference $\btheta_{t}^i - \btheta_{t-1}^i$ for this work. This is a flexible setup that admits several variants, e.g.,
		\begin{align}
		\| \btheta_{t}^i - \btheta_{t-1}^i \|_1 &= \sum_{j=1}^{d_{\theta}}|\btheta_{t,j}^i - \btheta_{t-1,j}^i|, \label{eqManhattannorm} \\
	\| \btheta_{t}^i - \btheta_{t-1}^i \|_2 &= 	\sqrt{ \sum_{j=1}^{d_\theta}(\btheta_{t,j}^i - \btheta_{t-1,j}^i)^2} \quad \text{and} \label{eqeuclideannorm}\\
		\| \btheta_{t}^i - \btheta_{t-1}^i \|_\infty &= \max_{1\leq j\leq d_{\theta}}|\btheta_{t,j}^i - \btheta_{t-1,j}^i|; \label{eqmaximumnorm}
		\end{align}
		i.e., the taxicab or Manhattan ($p=1$), the Euclidean norm ($p=2$) and the maximum norm ($p=\infty$) respectively.
} $\| \btheta_{t}^i - \btheta_{t-1}^i \|_p$  is computed and compared against a prescribed relative threshold $\lambda > 0$ in order to determine whether the prediction and update steps in the second layer of filters can be performed recursively or not. Specifically:
\begin{itemize}
	\item If $\| \btheta_{t}^i - \btheta_{t-1}^i \|_p  < \lambda\|\btheta_{t-1}^i\|_p$ is not satisfied for $\btheta_{t}^i$, the $i$-th filter runs from scratch following the scheme in Algorithm \ref{alEKF2ndlayer}.
	
	\item When $\|\btheta_{t}^i - \btheta_{t-1}^i\|_p < \lambda\|\btheta_{t-1}^i\|_p$ is satisfied for $\btheta_{t}^i$, only one prediction and update step (from time $t-1$ to time $t$) is needed. In particular, we make the approximation $p(\bx_{t-1} | \by_{1:t-1}, \btheta_{t}^i) \approx p(\bx_{t-1} | \by_{1:t-1}, \btheta_{t-1}^i)$.

\end{itemize}
In either case, we use $p(\bx_t | \by_{1:t-1}, \btheta_t^i)$ in order to compute $p(\by_t | \by_{1:t-1}, \btheta_t^i)$ as in step \ref{stepEKFlikelihood} of Algorithm \ref{alEKF2ndlayer}. Finally, we can compute the mean vector $\hat{\btheta}_t$ and the covariance matrix $\hat{\bC}_t^{\btheta,i}$ at time $t$ in step \ref{stepEstimatesall}, by using \eqref{eqParamEst} and \eqref{eqCovEst}. We prepare the new reference points $\btheta_{t+1}^i$ and their weights $w_{t+1}^i$ from $\mathcal{N}(\hat{\btheta}_t, \hat{\bC}_t^{\btheta,i})$ for the next time step.

\afterpage{ 

	\clearpage
\begin{algorithm} \caption{\label{alEKF2ndlayer} Extended Kalman filter conditional on $\btheta_t^i$, used in the second layer of the nested filter.}
	
	\textbf{Inputs}: 
	\begin{itemize}
		\item[-] Prior \gls{pdf} $p(\bx_0)$ and parameter vector $\btheta_t^i$.
		\item[-] State-space model described in equations \eqref{eqxdiscrete} and \eqref{eqydiscrete}. In particular, $f(\cdot)$ denotes the drift function in the state eq. \eqref{eqxdiscrete} and $g(\cdot)$ is the observation function in eq. \eqref{eqydiscrete}. The covariance of the state noise is denoted $\bV$ and the covariance of the observation noise is denoted $\bR$.
%		\begin{align}
%		\bx_t &= f(\bx_{t-1}, \btheta) + \bv_t,& \quad \bv_t \sim \mathcal{N}(\boldsymbol{0}, \bV), \\
%		\by_t &= g(\bx_t,\btheta) + \br_t,& \quad \br_t \sim \mathcal{N}(\boldsymbol{0}, \bR).
%		\end{align}
	\end{itemize}
	
	\textbf{Procedure}:
	\quad \begin{enumerate}	
		\item Initialization
		
		\begin{enumerate}
			\item Assume $p(\bx_0)$ is Gaussian with mean $\hat{\bx}_0$ and covariance $\hat{\bC}_0^{\bx}$, i.e., $p(\bx_0) \simeq \mathcal{N}(\bx_0 | \hat{\bx}_0, \hat{\bC}_0^{\bx})$.
		\end{enumerate}
		
		\item Sequential step, $t \ge 1$.	
		%In this step, it is assumed that the parameter and the state particle sets of the previous step is available for $n \ge 1$. This set have to be updated in the following way:
			\begin{enumerate}
				\item \textbf{Prediction step.} Compute 
					\begin{align}
					\tilde{\bx}_{t,\btheta_{t}^i} &= f(\hat{\bx}_{t-1,\btheta_{t}^i}, \btheta_{t}^i), \\
					\tilde{\bC}_{t,\btheta_{t}^i}^{\bx} &= \bJ_{f,\hat{\bx}_{t-1,\btheta_{t}^i}} \hat{\bC}_{t-1,\btheta_{t}^i}^{\bx} \bJ_{f,\hat{\bx}_{t-1,\btheta_{t}^i}}^\top + \bV,
					\end{align}
					where $\bJ_{f,\bx}$ is the Jacobian matrix of $f(\cdot)$ evaluated at $\hat{\bx}_{t-1,\btheta_{t}^i}$. \label{stepEKFpred}
			
				\item Approximate $p(\bx_{t} | \by_{1:t-1}, \btheta_{t}^i) \simeq \mathcal{N}(\bx_{t}| \tilde{\bx}_{t,\btheta_{t}^i}, \tilde{\bC}_{t,\btheta_{t}^i}^{\bx})$ and compute
					\begin{align}
				p(\by_t | \by_{1:t-1}, \btheta_{t}^i)& = \int p(\by_t | \bx_{t}, \btheta_t^i) p(\bx_{t} | \by_{1:t-1}, \btheta_{t}^i) d\bx_{t}\\
				& \simeq \int p(\by_t | \bx_{t}, \btheta_t^i) \mathcal{N}(\bx_t | \tilde{\bx}_{t, \btheta_t^i}, \tilde{\bC}_{t,\btheta_t^i}) d\bx_{t}.
				\end{align} \label{stepEKFlikelihood}			
			
%				\item Use the unscented transform or a Gaussian cubature rule to compute 
%				\begin{align}
%				&p(\by_t | \by_{1:t-1}, \btheta_{t}^i) \nonumber \\
%				&= \int p(\by_t | \bx_{t}, \btheta_t^i) p(\bx_{t} | \by_{1:t-1}, \btheta_{t}^i) d\bx_{t} \nonumber \\
%				&\simeq \int p(\by_t |\bx_{t}, \btheta_{t}^i) \mathcal{N}(\bx_{t}| \tilde{\bx}_{t,\btheta_{t}^i}, \tilde{\bC}_{t,\btheta_{t}^i}^{\bx}) d\bx_{t}. \nonumber
%				\end{align}	\label{stepEKFlikelihood}			
%				%Compute $p(\by_t | \by_{1:t-1}, {\btheta}_t^i)$ by approximating $\hat p(\bx_t | \by_{1:t-1}, {\btheta}_t^i ) = \mathcal{N}(\bx_{t}|\tilde{\bx}_{t,\btheta_{t}^i}, \tilde{\bC}_{t,\btheta_{t}^i}^{\bx})$. 
				
				\item \textbf{Update step.} Compute
				\begin{eqnarray}
				\hat{\bx}_{t,\btheta_{t}^i} &=& \tilde{\bx}_{t,\btheta_{t}^i} + \bK_t (\by_t - g(\tilde{\bx}_{t,\btheta_{t}^i}, \btheta_{t}^i)), \\
				\hat{\bC}_{t,\btheta_{t}^i}^{\bx} &=& (\boldsymbol{I}_{d_x} - \bK_t \bJ_g) 	\tilde{\bC}_{t,\btheta_{t}^i}^{\bx} , \\
				\bK_t &=& \tilde{\bC}_{t,\btheta_{t}^i}^{\bx} \bJ_{g,\tilde{\bx}_{t,\btheta_{t}^i}}^\top (\bJ_{g,\tilde{\bx}_{t,\btheta_{t}^i}} \tilde{\bC}_{t,\btheta_{t}^i}^{\bx} \bJ_{g,\tilde{\bx}_{t,\btheta_{t}^i}}^\top + \bR), \nonumber
				\end{eqnarray}
				where $\bJ_{g,\bx}$ is the Jacobian matrix of $g(\cdot)$ evaluated at $\tilde{\bx}_{t,\btheta_{t}^i}$. Approximate $p(\bx_t | \by_{1:t}, {\btheta}) \simeq \mathcal{N}(\bx_{t}|\hat{\bx}_{t,\btheta_{t}^i}, \hat{\bC}_{t,\btheta_{t}^i}^{\bx})$.
				
			\end{enumerate}

	\end{enumerate}

	\textbf{Outputs}: $\hat{\bx}_{t,\btheta_{t}^i}$, $\hat{\bC}_{t,\btheta_{t}^i}^{\bx}$ and $p(\by_t | \by_{1:t-1}, {\btheta}_t^i)$.
\end{algorithm} 	

\thispagestyle{empty}
\clearpage 
}

\subsection{State tracking} \label{ssStatetracking}

We can take advantage of the filters in the second layer in order to provide state estimates as well. Let us write the expectation of $\bx_{t}$ as
\begin{equation}
\mathbb{E}[\bx_{t}|\by_{1:t}] = \int_{{\btheta}} \Big[ \int_{\mathcal{X}} \bx_{t} p(\bx_{t} | \btheta, \by_{1:t}) d\bx_{t} \Big] p(\btheta | \by_{1:t}) d\btheta,
\end{equation}
where the integral in square brackets can be approximated by the $M$ Gaussian filters of the second layer. In this case, we assume they are the \glspl{ekf} of Algorithm \ref{alEKF2ndlayer} conditional on $\btheta = \btheta_{t}^i$. This yields a Gaussian approximation $p(\bx_{t}|\btheta_{t}^i, \by_{1:t}) \simeq \mathcal{N}(\bx_{t}|\hat{\bx}_{t,\btheta_{t}^i}, \hat{\bC}_{t,\btheta_{t}^i})$, where 
\begin{align}
\hat{\bx}_{t,\btheta_{t}^i} &\simeq \mathbb{E}[\bx_{t}|\btheta_{t}^i, \by_{1:t}] \quad \text{and} \\
\hat{\bC}_{t,\btheta_{t}^i}^{\bx} &\simeq \mathbb{E}[(\bx_{t} - \hat{\bx}_{t,\btheta_{t}^i} )(\bx_{t} - \hat{\bx}_{t,\btheta_{t}^i} )^\top | \by_{1:t}, \btheta_{t}^i]. 
\end{align}
Then, a Gaussian approximation $p(\bx_{t} | \by_{1:t}) \simeq \mathcal{N}(\bx_{t} | \hat{\bx}_t, \hat{\bC}_t^{\bx})$ can be constructed, where
\begin{align}
\hat{\bx}_{t} &\simeq \sum_{i=1}^{M} \hat{\bx}_{t,\btheta^i_t} \frac{p(\by_t | \by_{1:t-1}, {\btheta}_t^{i})}{p(\by_t | \by_{1:t-1})} w_t^{i} \quad \text{and} \label{eqStateEst}\\
\hat{\bC}^{\bx}_t &\simeq \sum_{i=1}^{M} (\hat{\bx}_{t,\btheta^i_t} - \hat{\bx}_t) (\hat{\bx}_{t,\btheta^i_t} - \hat{\bx}_t)^{\top} \frac{p(\by_t | \by_{1:t-1}, {\btheta}_t^{i})}{p(\by_t| \by_{1:t-1})} w_t^{i}. \label{eqCovStateEst}
\end{align}

%\textcolor{red}{Explain lambda. Euclidean distance vs max. Here?}
\subsection{Continuity of the conditional filter \gls{pdf}}

The key to keep Algorithm \ref{alNGFRecursive} recursive is the test in step \ref{stepRecursivity}, which sets
\begin{equation}
p(\bx_{t-1}|\by_{1:t-1}, \btheta_{t}^i) \simeq \mathcal{N}(\bx_{t-1} |\hat{\bx}_{t-1,\btheta_{t-1}^i}, \hat{\bC}^{\bx}_{t-1,\btheta_{t-1}^i})
\end{equation}
when $\frac{\|\btheta_{t}^i - \btheta_{t-1}^i\|_p}{\|\btheta_{t-1}^i\|_p} < \lambda$ for some prescribed threshold $\lambda > 0$. This step relies on the assumption that $p(\bx_{t-1}|\by_{1:t-1}, \btheta) \approx p(\bx_{t-1}| \by_{1:t-1},\btheta')$ when $\btheta\approx \btheta'$, i.e., we are assuming that the conditional filtering \gls{pdf} $p(\bx_t | \by_{1:t}, \btheta)$ is a continuous function of the parameter $\btheta$. In this section we state sufficient conditions for the conditional filter $p(\bx_t|\by_{1:t},\btheta)$ to be Lipschitz-continuous.

For conciseness, let us denote
\begin{eqnarray}
\pi_t(\bx_t|\btheta) &\coloneqq& p(\bx_t | \by_{1:t}, \btheta), \\
\xi_t(\bx_t|\btheta) &\coloneqq& p(\bx_t | \by_{1:t-1}, \btheta), \quad \text{and}\\
\eta_t(\by_t|\btheta) &\coloneqq& p(\by_t| \by_{1:t-1},\btheta).
\end{eqnarray}

Hereafter we assume that the observation sequence $\{\by_t,t \ge 1\}$ is arbitrary but fixed (i.e., deterministic). Additionally, we impose the following regularity assumptions:

\begin{Assumption} \label{as1}
	The conditional \glspl{pdf} $\pi_t(\bx_t|\btheta)$, $\xi_t(\bx_t|\btheta)$ and $\eta_t(\by_t|\btheta)$ exist for every $t\ge1$, every $\bx_t \in \mathbb{R}^{d_x}$ and every parameter vector $\btheta \in \Theta \subseteq \mathbb{R}^{d_\theta}$, where $\Theta$ denotes the parameter space.
\end{Assumption}

\begin{Assumption} \label{as2}
	The transition \gls{pdf} $p(\bx_t|\bx_{t-1},\btheta)$ is Lipschitz \gls{wrt} $\btheta$, i.e., there exists a constant $0<L<\infty$ such that 
	\begin{equation}
	\sup_{\bx_{t-1}\in\mathbb{R}^{d_x}} \int |p(\bx_t|\bx_{t-1},\btheta) - p(\bx_t|\bx_{t-1},\btheta')|d\bx_t < L \|\btheta-\btheta'\|
	\end{equation}
	for every $t\ge 1$ and every pair $(\btheta,\btheta')\in\Theta\times\Theta$.
\end{Assumption}

\begin{Remark}
	In Assumption \ref{as2}, we denote $\|\btheta-\btheta'\|=\sqrt{\sum_{i=1}^{d_{\theta}}(\theta_i-\theta_i')^2}$, the Euclidean distance between $\btheta$ and $\btheta'$.
\end{Remark}

\begin{Assumption} \label{as3}
	The conditional \glspl{pdf} $p(\by_t|\bx_t,\btheta)$ are strictly positive and uniformly Lipschitz \gls{wrt} $\btheta$. In particular, $p(\by_t|\bx_t,\btheta) > 0$ and
	\begin{equation}
	\sup_{\bx_t \in \mathbb{R}^{d_x}} \frac{|p(\by_t|\bx_t,\btheta)- p(\by_t|\bx_t,\btheta')|}{\eta_t(\by_t|\btheta)} < G_t \|\btheta-\btheta'\|
	\end{equation}
	for some positive $G_t < \infty$.
\end{Assumption}

\begin{Assumption} \label{as4}
	The ratio $\frac{p(\by_t|\bx_t,\btheta)}{\eta_t(\by_t|\btheta)}$ is bounded. Specifically, there exist finite constants $0 < M_t < \infty$ such that
	\begin{equation}
	\sup_{\substack{\btheta\in\Theta \\ \bx_{t-1}\in\mathbb{R}^{d_x}}} \frac{p(\by_t|\bx_t,\btheta)}{\eta_t(\by_t|\btheta)} < M_t.
	\end{equation}
\end{Assumption}

Assumptions \ref{as1} and \ref{as2} are rather mild and easy to check for a given state-space model. Assumptions \ref{as3} and \ref{as4}, on the other hand, may be restrictive in some problems. We note, however, that for fixed $\by_t$, $t\ge1$, and a compact parameter support $\Theta\subset\mathbb{R}^{d_\theta}$, the factor $\eta_t(\by_t|\btheta)$ can often be bounded away from $0$, while $p(\by_t|\bx_t,\btheta)$ is typically upper bounded. In any case, Assumptions \ref{as1}-\ref{as4} lead to the result below regarding the continuity of the filter, $\pi_t(\bx_t|\btheta)$, and predictive, $\xi_t(\bx_t|\btheta)$, \glspl{pdf} \gls{wrt} the parameter vector $\btheta$.

\begin{Proposition} \label{prop1}
	: If Assumptions \ref{as1} to \ref{as4} hold, there exist sequences of finite constants $\tilde{L}_t$ and $L_t$ such that, for $t \ge 1$,
	\begin{eqnarray}
	\int |\xi_t(\bx_t|\btheta)- \xi_t(\bx_t|\btheta')|d\bx_t &\le& \tilde{L}_t \|\btheta-\btheta'\|, \quad \text{and}\\
	\int |\pi_t(\bx_t|\btheta)- \pi_t(\bx_t|\btheta')|d\bx_t&\le& L_t \|\btheta-\btheta'\|.
	\end{eqnarray}
\end{Proposition}

See \ref{approof} for a proof.

%$d(\btheta_{t}^i , \btheta_{t-1}^i) = \max_{1\leq j\leq d_{\theta}}|\btheta_{t,j}^i - \btheta_{t-1,j}^i|$; the taxicab distance, $d(\btheta_{t}^i , \btheta_{t-1}^i) = \sum_{j=1}^{d_{\theta}}|\btheta_{t,j}^i - \btheta_{t-1,j}^i|$; etc. 
%We adopt a relative threshold in step \ref{stepThreshold} by not only introducing $\lambda$ but also taking into account the value of the estimated parameters in the previous time step, $\btheta_{t-1}^i$. Therefore, with the comparison $\parallel \btheta_{t}^i - \btheta_{t-1}^i \parallel_p < \lambda \parallel \btheta_{t-1}^i \parallel_p$, this step adapts to the estimate $\btheta_{t-1}^i$ currently available.

%\afterpage{ 	\clearpage
\begin{algorithm} \caption{Recursive nested Gaussian filters.\label{alNGFRecursive} }
	\textbf{Inputs}: 
	\begin{itemize}
		\item[-] Prior pdfs $p(\bx_0)$ and $p(\btheta)$.
		\item[-] A fixed threshold $\lambda > 0$.
	\end{itemize}	
	\textbf{Procedure}:
	\quad \begin{enumerate}	
		\item Initialization		
		\begin{enumerate}
			\item Generate $M$ reference points, $\btheta^i_1$, for $p(\btheta) \simeq \mathcal{N}(\btheta_0, \bC_0^{\btheta})$, $i = 1,\ldots,M$, with weights $w_1^i$. \label{stepInitRefPoint}
			\item If $p(\bx_0) = \mathcal{N} (\bx_0 | \hat{\bx}_0, \bC_0^{\bx})$, then set $\hat{\bx}_0^i = \hat{\bx}_0$ and $\hat{\bC}_0^{\bx,i} = \bC_0^{\bx}$ for $i = 1,\ldots,M$. \label{stepInitX}
		\end{enumerate}
		\item Sequential step, $t \ge 1$.	
		%In this step, it is assumed that the parameter and the state particle sets of the previous step is available for $n \ge 1$. This set have to be updated in the following way:
		\begin{enumerate}
			\item For $i=1,\ldots,M$: \label{stepRecursivity}
			\begin{enumerate}
				\item If $\|\btheta_{t}^i-\btheta_{t-1}^i\|_p<\lambda\|\btheta_{t-1}^i\|_p$, then compute $p(\bx_t | \by_{1:t-1}, {\btheta}_t^i )$ from $p(\bx_{t-1} | \by_{1:t-1}, \btheta_{t}^i) \simeq p(\bx_{t-1} | \by_{1:t-1}, \btheta_{t-1}^i)$, where $p(\bx_{t-1} | \by_{1:t-1}, \btheta_{t-1}^i) \simeq \mathcal{N}(\bx_{t-1}|\hat{\bx}_{t-1,\btheta_{t-1}^i}, \hat{\bC}_{t-1,\btheta_{t-1}^i}^{\bx})$.
				Else, approximate $p(\bx_t | \by_{1:t-1}, {\btheta}_t^i )$ from the prior $p(\bx_0)$. \label{stepThreshold}
				\item Use $p(\bx_t | \by_{1:t-1}, \btheta_t^i)$ to compute $p(\by_t | \by_{1:t-1}, {\btheta}_t^{i})$. \label{stepOutput2ndlayer}
			\end{enumerate}
			\item Compute $\hat{\btheta}_t$, $\hat{\bC}^{\btheta}_t$, $\hat{\bx}_t$ and $\hat{\bC}^{\bx}_t$ from \eqref{eqParamEst}, \eqref{eqCovEst}, \eqref{eqStateEst} and \eqref{eqCovStateEst}, respectively. \label{stepEstimatesall}
			\item Generate reference points $\btheta^i_{t+1}$ and weights $w_{t+1}^i$ from $\hat{\btheta}_t$ and $\hat{\bC}^{\btheta}_{t}$ for $i = 1,\ldots,M$. \label{stepUpdateRefPoint}
			
		\end{enumerate}
		
	\end{enumerate}
	
	\textbf{Outputs}: $\hat{\bx}_t$, $\hat{\btheta}_t$, $\hat{\bC}_t^{\bx}$ and $\hat{\bC}_t^{\btheta}$.
\end{algorithm}
%\thispagestyle{empty}
%\clearpage 
%} 

\section{Example} \label{sExample}

\subsection{Stochastic Lorenz 63 model} \label{ssStochasticL63model}

Consider the 3-dimensional continuous-time stochastic process $\bx(\tau) = [x_1(\tau), x_2(\tau), x_3(\tau)]^\top$, for $\tau \in (0,\infty)$, taking values on $\real^3$, whose dynamics are described by the system of stochastic differential equations (SDEs)
\begin{align}
	dx_{1} &= -S (x_{1} - x_{2}) + \sigma dv_{1}, \label{eqlorenz63sde1}\\
	dx_{2} &= Rx_{1} - x_{2} - x_{1}x_{3} + \sigma dv_{2}, \label{eqlorenz63sde2}\\
	dx_{3} &= x_{1}x_{2} - Bx_{3} + \sigma dv_{3}, \label{eqlorenz63sde3}
\end{align}
where the $v_{i}$'s are independent 1-dimensional Wiener processes, $\sigma > 0$ is a known scale parameter and $S,R,B \in \real$ are unknown static model parameters. Using the Euler-Maruyama scheme in order to integrate the SDEs \eqref{eqlorenz63sde1}--\eqref{eqlorenz63sde3}, it is straightforward to convert them into the discrete-time state equation
%write the state-space model as
\begin{equation}
\bx_{t+1} = f_{\Delta}(\bx_{t},\btheta) + \sqrt{\Delta} \bv_t, \quad t = 1,2,\ldots\label{eql63}
\end{equation}
where $f_{\Delta} : \real^{d_x} \times \real^{d_\theta} \to \real^{d_x}$ ($d_x=d_\theta = 3$) is the function defined by
%can be expressed as
\begin{align}
 f_{1,\Delta}(\bx_{t},\btheta) &= x_{1,t} - \Delta S (x_{1,t} - x_{2,t}), \nonumber  \\  
 f_{2,\Delta}(\bx_{t},\btheta) &= x_{2,t} + \Delta [(R-x_{3,t})x_{1,t} - x_{2,t}], \nonumber \\  
 f_{3,\Delta}(\bx_{t},\btheta) &= x_{3,t} + \Delta (x_{1,t}x_{2,t} - Bx_{3,t}),\nonumber 
\end{align}
 $\Delta$ is the integration time-step, $\btheta = (S,R,B)^\top$ is the $3\times 1$ vector of unknown parameters and $\bv_t$ is a sequence of $3$-dimensional Gaussian independent random vectors with zero mean and covariance matrix $\sigma_x^2 \boldsymbol{I}_{3}$ (with $\boldsymbol{I}_d$ denoting the $d \times d$ identity matrix). Hence, the state transition density $p(\bx_t | \bx_{t-1}, \btheta)$ is Gaussian and can be written down as $p(\bx_t | \bx_{t-1}, \btheta) = \mathcal{N} (\bx_{t} | f_{\Delta}(\bx_{t-1},\btheta), \sigma^2 \Delta \boldsymbol{I}_{d_x})$. This function is Lipschitz on $\btheta$.

In order to complete the specification of a state space model, we need to characterize the observations. For our simulation setup we assume linear observations of the form 
%We assume the observation equation
\begin{align}
	\by_t = k_o \begin{bmatrix}
	x_{1,t} \\
	x_{3,t}  \\
	\end{bmatrix}
	+ \br_t, \label{eqObs63}
\end{align}
where $k_o$ is a fixed parameter and $\br_t \sim \mathcal{N}(\br_t | \boldsymbol{0}, \sigma_{y}^2 \boldsymbol{I}_2)$ is a $2$-dimensional additive noise with zero mean and covariance function $\sigma_{y}^2 \boldsymbol{I}_2$. Therefore, the conditional part of the observations (and hence the likelihood function) is also Gaussian and can be written as $p(\by_t|\bx_t) = \mathcal{N}(\by_t | \boldsymbol{G} \bx_t, \sigma_{y}^2 \boldsymbol{I}_2)$, where $\boldsymbol{G} = \begin{bmatrix}
k_o & 0 & 0 \\
0 & 0 & k_o  \\
\end{bmatrix}$ is the observation matrix. This function has a finite upper bound independent of $\btheta$.

Observations are not collected at every time $t$. Instead we assume that an observation vector is received every $M_o$ steps of the state equation.
%Gaussian random variable with covariance matrix $\sigma_y^2 \boldsymbol{I}_{2}$.

\subsection{Simulation setup} \label{ssSimulationSetup}

For our computer experiments we have used the stochastic Lorenz 63 model outlined in \eqref{eql63} and \eqref{eqObs63} in order to generate signals $\bx_t$ and $\by_t, t = \{0,1,\ldots\}$, used as the ground truth and the data, respectively, for the assessment of the algorithm. We integrate the model with the step-size $\Delta = 2 \times 10^{-4}$ continuous-time units.  The true parameters for the generation of the signal and data are $S = 10$, $R = 28$ and $B=\frac{8}{3}$ (which yield underlying chaotic dynamics); while the initial state is Gaussian with mean\footnote{The initial vector $\hat{\bx}_0$ is taken from a deterministic realization of the Lorenz 63 model.} $\hat{\bx}_0 = [-6, -5.5, -24.5]^\top $ and covariance matrix $\boldsymbol{I}_3$, i.e., $p(\bx_0| \hat{\bx}_0, \boldsymbol{I}_3)$ . The noise scale factors, $\sigma^2 = 0.1$ and $\sigma_y^2 = 1$, are assumed known. 

For the estimation task we use Algorithm \ref{alNGFRecursive}. We assume a Gaussian prior distribution for the unknown parameters, namely $p(\btheta) = \mathcal{N}(\btheta|\boldsymbol{\mu}_\theta,\boldsymbol{I}_3)$, where the a priori mean $\boldsymbol{\mu}_\theta$ is drawn at random from a uniform distribution $\mathcal{U}(\btheta_\star - \boldsymbol{\epsilon},\btheta_\star + \boldsymbol{\epsilon})$ for each independent simulation. $\btheta_\star = [10, 28, \frac{8}{3}]^\top $ are the true parameter vector and the offset vector is $\boldsymbol{\epsilon} = [3, 1, 0.5]^\top$. The algorithm does not collect an observation at every time step, but every $M_o=5$ discrete-time steps ($10^{-3}$ continuous-time units). Hence, the prediction step of the state variables at the second layer of nested filter corresponds to $M_o=5$ discrete-time steps of the Euler scheme. When an observation $\by_t$ (at time $t=kM_o$ ($k \in \mathbb{N}$)) arrives, both the state and parameter distributions are updated. The length of each simulation runs is $T=40$ continuous-time units ($2 \times 10^{5}$ discrete-time steps of the state equation \eqref{eql63}).

%We assume a threshold $\lambda = 5 \times 10^{-3}$. Every simulation runs until $t = T = 10^5$.

We have assessed the ability of several Bayesian computation algorithms to jointly track the state $\bx_t$ and estimate the parameters $\btheta=(S,R,B)^\top$ of this model. To be specific, we have coded and run the following schemes:
\begin{itemize}
	\item The proposed Algorithm \ref{alNGFRecursive} using an \gls{ukf} in the first layer and a bank of \glspl{ekf} in the second layer.
	\item A \gls{ukf} \cite{Julier00} algorithm with state augmentation \cite{Lourens12,Hassanzadeh08} where the parameters are added to the state vector.
	\item An \gls{enkf} \cite{Evensen03} algorithm with state augmentation as well.
	\item A \gls{nhf} \cite{Perez-Vieites18} with a \gls{smc} algorithm in the first layer and a bank of \glspl{ekf} in the second layer.
\end{itemize}

The accuracy of the various algorithms is compared in terms of the normalized mean square error (NMSE) of the predictor of the state and the predictor of the parameters. We assess the empirical NMSE resulting directly from the simulations, namely,
	\begin{equation}
	\text{NMSE}_{x,t} = \frac{\| \bx_t - \hat{\bx}_t \|^2}{\| \bx_t \|^2}, \quad 
		\text{NMSE}_{\btheta,t} = \frac{\| \btheta_t - \hat{\btheta}_t \|^2}{\| \btheta_t \|^2}, \label{eqNMSEtt}
	\end{equation}
	as well as the averages $\text{NMSE}_{x} = \frac{1}{T} \sum_{t=0}^{T-1}  \text{NMSE}_{x,t}$ and $\text{NMSE}_{\btheta} = \frac{1}{T} \sum_{t=0}^{T-1}  \text{NMSE}_{\btheta,t}$.
	%being NMSE$_x$ and NMSE$_{\btheta}$ the average of NMSE$_{x,t}$ and NMSE$_{\btheta,t}$ over time.

	\subsection{Numerical results} \label{ssNumericalResults}
	
	In the first computer experiments we study the choice of norm $\|\btheta_{t}^i - \btheta_{t-1}^i\|_p$ in step \ref{stepThreshold} of Algorithm \ref{alNGFRecursive}. Specifically, we have considered a setup where the model parameters $\btheta = (S,R,B)^\top$ are assumed known and the goal is to track the state $\bx_t$ using an \gls{ekf}. We first generate a sequence of observations $\by_{1:T}$ from the model with parameters $\btheta=(10,28,\frac{8}{3})^\top$. Then, for this sequence, the \gls{ekf} runs with a perturbed set of parameters of the form $\btheta'=\btheta+\epsilon$, where $\epsilon \sim \mathcal{N}(0,\sigma_e^2)$ is a zero-mean Gaussian perturbation. We carry out 100 independent simulations for each value of $\sigma_e^2$ for $\sigma_e^2=\{10^{-1},10^{-2}, 10^{-3},10^{-4},10^{-5}\}$.
	
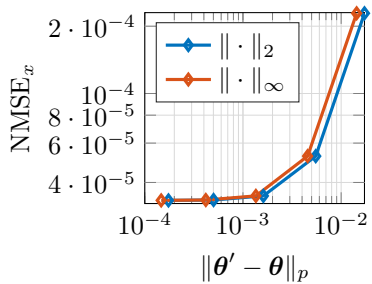
\begin{figure}[htb]
	\centering
\definecolor{mycolor1}{rgb}{0.00000,0.44700,0.74100}%
\definecolor{mycolor2}{rgb}{0.85000,0.32500,0.09800}%
\begin{tikzpicture}

\begin{axis}[%
width=1.15in,
height=1in,
at={(1.011in,0.642in)},
scale only axis,
xmode=log,
xmin=0.0001,
xmax=0.017282171970154,
xminorticks=true,
xlabel={$\|\btheta' - \btheta\|_p$},
ymode=log,
ymin=3.22e-05,
ymax=0.00023,
yminorticks=true,
ylabel={$\text{NMSE}_{x}$},
ylabel style= {yshift=0.4cm},
extra y ticks={4e-05,6e-05,8e-05, 2e-4},
grid=both,
grid style={line width=.1pt, draw=gray!30},
axis background/.style={fill=white},
legend style={at={(0.05,0.95)},anchor=north west,legend cell align=left,align=left,draw=white!15!black}
]
\addplot [color=mycolor1,solid,line width=1.2pt,mark=diamond,mark options={solid}]
  table[row sep=crcr]{%
0	3.32396576199244e-05\\
0.000174404422965432	3.3260854116603e-05\\
0.000502016214064147	3.33908347815175e-05\\
0.00161394417563293	3.48381197025377e-05\\
0.00550832641885537	5.2459084666672e-05\\
0.017282171970154	0.000228443438866359\\
};
\addlegendentry{$\| \cdot \|_{2}$};

\addplot [color=mycolor2,solid,line width=1.2pt,mark=diamond,mark options={solid}]
  table[row sep=crcr]{%
0	3.32396576199244e-05\\
0.000147017867330047	3.3260854116603e-05\\
0.00041851699264794	3.33908347815175e-05\\
0.00135378489208392	3.48381197025377e-05\\
0.00455539649780931	5.2459084666672e-05\\
0.014380476673464	0.000228443438866359\\
};
\addlegendentry{$\| \cdot \|_{\infty}$};

\end{axis}

%\begin{axis}[%
%width=7.778in,
%height=5.833in,
%at={(0in,0in)},
%scale only axis,
%xmin=0,
%xmax=1,
%ymin=0,
%ymax=1,
%hide axis,
%axis x line*=bottom,
%axis y line*=left,
%legend style={legend cell align=left,align=left,draw=white!15!black}
%]
%\end{axis}
\end{tikzpicture}%
	\caption{EKF performance with know parameters, $\btheta'$, that are obtained by modifying the true parameters, $\btheta$. In the abscissa axis, we represent the average distance of the simulation runs to the ground truh.
			%being $\mathbb{E}[d]= \mathbb{E}[\parallel \btheta - \btheta' \parallel]$ for the euclidean distance and $\mathbb{E}[d]= \mathbb{E}[\max_{1\leq j\leq d_{\theta}} |\btheta_j - \btheta'_j |]$ for the distance of the maximum. 
		}
	\label{figMSEEdistEKFs}
\end{figure}

	Figure \ref{figMSEEdistEKFs} summarizes the outcome of this experiment. In particular, it displays the NMSE in the tracking of $\bx_t$, averaged over all 100 simulation runs, versus the average norms $\|\btheta-\btheta'\|_2$ and $\|\btheta-\btheta'\|_\infty$. The plot illustrates that:
	\begin{enumerate}[i.]
		\item The NMSE$_x$ is a continuous magnitude \gls{wrt} the perturbation $\|\btheta - \btheta'\|$, both with Euclidean or maximum norms. The NMSE$_x$ remains below $10^{-4}$ when $\|\btheta - \btheta'\|$ is approximately below $10^{-2}$.
		\item The NMSE$_x$ is slightly higher when the parameter perturbation is given in terms of the norm $\|\btheta - \btheta'\|_\infty$.
	\end{enumerate}

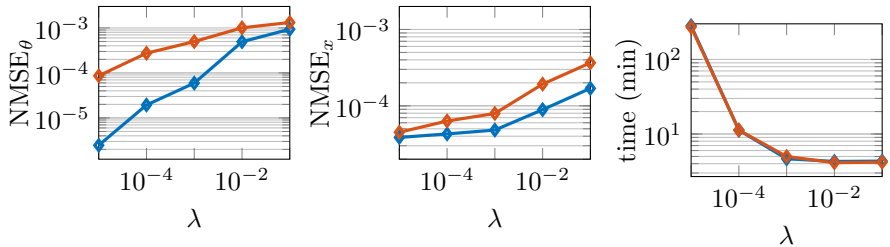
\begin{figure}[htb]
	\centering
	\begin{subfigure}{0.32\linewidth}
		\centering
\definecolor{mycolor1}{rgb}{0.00000,0.44700,0.74100}%
\definecolor{mycolor2}{rgb}{0.85000,0.32500,0.09800}%
\begin{tikzpicture}

\begin{axis}[%
width=1in,
height=0.8in,
at={(1.011in,0.642in)},
scale only axis,
xmode=log,
xmin=1e-05,
xmax=0.1,
xminorticks=false,
xlabel={$\lambda$},
ymode=log,
ymin=0e-7,
ymax=0.003,
ylabel style= {yshift=-0.2cm},
yminorticks=true,
yminorgrids,
ymajorgrids,
xtick={0.00001,0.0001,0.001,0.01,0.1},
xticklabels={,10$^{-4}$,,10$^{-2}$,},
ylabel={$\text{NMSE}_{\theta}$},
axis background/.style={fill=white},
%legend pos=south east,
legend style={at={(1,0.42)},legend cell align=left,align=left,draw=white!15!black}
]
\addplot [color=mycolor1,solid,line width=1.2pt,mark=diamond,mark options={solid}]
  table[row sep=crcr]{%
0.1	0.000934150112355829\\
0.01	0.000492962723328069\\
0.001	5.93530520301606e-05\\
0.0001	1.93813219854081e-05\\
1e-05	2.46243377075387e-06\\
};
%\addlegendentry{$\| \cdot \|_{2}$};

\addplot [color=mycolor2,solid,line width=1.2pt,mark=diamond,mark options={solid}]
  table[row sep=crcr]{%
0.1	0.00132162465344314\\
0.01	0.00100625888837254\\
0.001	0.000497491261963896\\
0.0001	0.000274238476420422\\
1e-05	8.64411891147016e-05\\
};
%\addlegendentry{$\| \cdot \|_{\infty}$};

\end{axis}
\end{tikzpicture}%
		\caption{NMSE$_{\btheta}$ for different values of $\lambda$.}
		%\caption{}
		\label{figmsethetalambda}
	\end{subfigure}
	%\hfill
	\begin{subfigure}{0.32\linewidth}
		\centering
\definecolor{mycolor1}{rgb}{0.00000,0.44700,0.74100}%
\definecolor{mycolor2}{rgb}{0.85000,0.32500,0.09800}%
\begin{tikzpicture}

\begin{axis}[%
width=1in,
height=0.8in,
at={(1.011in,0.642in)},
scale only axis,
xmode=log,
xmin=1e-05,
xmax=0.1,
xminorticks=false,
xlabel={$\lambda$},
ymode=log,
ymin=0.00002,
ymax=0.002,
yminorticks=true,
yminorgrids,
ymajorgrids,
xtick={0.00001,0.0001,0.001,0.01,0.1},
xticklabels={,10$^{-4}$,,10$^{-2}$,},
ylabel style= {yshift=-0.2cm},
ylabel={$\text{NMSE}_{x}$},
axis background/.style={fill=white},
legend style={at={(0.7,1)},legend cell align=left,align=left,draw=white!15!black}
]
\addplot [color=mycolor1,solid,line width=1.2pt,mark=diamond,mark options={solid}]
  table[row sep=crcr]{%
0.1	0.000169215717557847\\
0.01	8.9017542979738e-05\\
0.001	4.81284190586957e-05\\
0.0001	4.28553985463092e-05\\
1e-05	3.85643399687197e-05\\
};
%\addlegendentry{$\| \cdot \|_{2}$};

\addplot [color=mycolor2,solid,line width=1.2pt,mark=diamond,mark options={solid}]
  table[row sep=crcr]{%
0.1	0.000365735640156085\\
0.01	0.000192349041509533\\
0.001	7.93350990939508e-05\\
0.0001	6.32053522600013e-05\\
1e-05	4.50281792495497e-05\\
};
%\addlegendentry{$\| \cdot \|_{\infty}$};

\end{axis}
\end{tikzpicture}%
		\caption{NMSE$_x$ for different values of $\lambda$.}
		%\caption{}
		\label{figmsexlambda}
	\end{subfigure}
	\begin{subfigure}{0.32\linewidth}
		\centering
\definecolor{mycolor1}{rgb}{0.00000,0.44700,0.74100}%
\definecolor{mycolor2}{rgb}{0.85000,0.32500,0.09800}%
\begin{tikzpicture}

\begin{axis}[%
width=1in,
height=0.8in,
at={(1.011in,0.642in)},
scale only axis,
xmode=log,
xmin=1e-05,
xmax=0.1,
xminorticks=false,
xlabel={$\lambda$},
ymode=log,
ymin=0,
ymax=300,
yminorticks=true,
yminorgrids,
ymajorgrids,
xtick={0.00001,0.0001,0.001,0.01,0.1},
xticklabels={,10$^{-4}$,,10$^{-2}$,},
ylabel={time (min)},
ylabel style= {yshift=-0.4cm},
axis background/.style={fill=white},
legend style={legend cell align=left,align=left,draw=white!15!black}
]
\addplot [color=mycolor1,solid,line width=1.2pt,mark=diamond,mark options={solid}]
  table[row sep=crcr]{%
0.1	4.33016501541667\\
0.01	4.285195738125\\
0.001	4.65485955229167\\
0.0001	11.2490454860417\\
1e-05	282.421889342857\\
};
%\addlegendentry{$\| \cdot \|_{2}$};

\addplot [color=mycolor2,solid,line width=1.2pt,mark=diamond,mark options={solid}]
  table[row sep=crcr]{%
0.1	4.18527528208333\\
0.01	4.12683775666667\\
0.001	4.98802997520833\\
0.0001	11.2608197741667\\
1e-05	271.40196167381\\
};
%\addlegendentry{$\| \cdot \|_{\infty}$};

\end{axis}
\end{tikzpicture}%
		\caption{Average simulation runtime.}
		%\caption{}
		\label{figtimelambda}
	\end{subfigure}
	%	\hfill
	%	\begin{subfigure}{0.49\linewidth}
	%		\centering
	%	\includegraphics[width=0.99\linewidth]{figures2/timelambda_less.eps}
	%		\caption{Average time of simulation.}
	%	\end{subfigure}
	\caption{The average NMSE and average time of simulation in minutes over $80$ simulation runs for different values of $\lambda$, using the Euclidean norm $\|\cdot\|_2$ (in blue) and the maximum norm $\|\cdot\|_\infty$ (in red). }
	\label{figlambda}
\end{figure}

 In a second experiment, Figure \ref{figlambda} shows the results of using Algorithm \ref{alNGFRecursive} with both $\| \cdot \|_2$ and $\| \cdot \|_\infty$ norms for several values of $\lambda$. Again, each point of the graphs represents the average of 80 independent simulation runs. We display NMSE$_{\btheta}$, NMSE$_{\bx}$ and run-times in minutes\footnote{The algorithms have been coded in MATLAB R2017a and run on a computer with 128 GB of DRAM and equipped with two Intel Xeon Gold 5115 10-Core CPU processors (running at
 2.40 GHz).} in Figures \ref{figmsethetalambda}, \ref{figmsexlambda} and \ref{figtimelambda}, respectively.  In Figure \ref{figmsethetalambda}, we see that NMSE$_{\btheta}$ increases with $\lambda$. This is as expected be cause the larger $\lambda$ the worser the approximation $p(\bx_{t-1}| \by_{1:t-1}, \btheta_{t}^i) \sim p(\bx_{t-1} | \by_{1:t-1}, \btheta_{t-1}^i)$. We also see that the Euclidean norm $\| \cdot \|_2$ yields a smaller error. However, in the results obtained for NMSE$_x$ in Figure \ref{figmsexlambda}, we observe that  below $\lambda=10^{-3}$ there is almost no improvement in the error, and the curve is similar to the one in Figure \ref{figMSEEdistEKFs}. Finally, Figure \ref{figtimelambda} shows that the runtime of the nested filtering Algorithm \ref{alNGFRecursive} increases significantly when $\lambda < 10^{-3}$ (because the algorithm takes longer to become strictly recursive). Therefore, we set $\lambda=10^{-3}$ in the following experiments as it appears to yield a good trade-off between accuracy and computational cost.

\begin{figure}[htb]
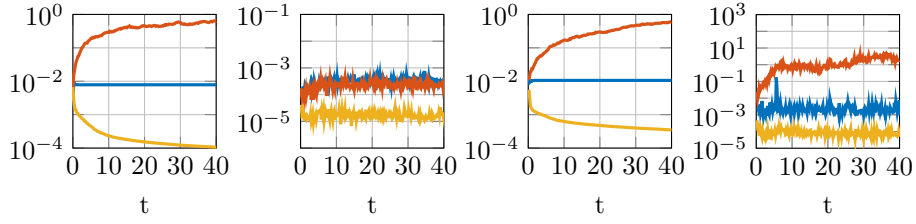

	\centering
	\begin{subfigure}{0.24\linewidth}
		\centering
		\input{EnKFvsAlg_obsall_MSEthetav3.tex}
		\caption{NMSE$_{\btheta,t}$ observing $[x_1, x_2, x_3]^\top$.}
		\label{fignmsetheta3states}
	\end{subfigure}
	%\hfill
	\begin{subfigure}{0.24\linewidth}
		\centering
		\input{EnKFvsAlg_obsall_MSExv3.tex}
		\caption{NMSE$_{x,t}$ observing $[x_1, x_2, x_3]^\top$. }
		\label{fignmsex3states}
	\end{subfigure}
	\begin{subfigure}{0.24\linewidth}
		\centering
		\input{EnKFvsAlg_obsxandz_MSEthetav3.tex}
		\caption{NMSE$_{\btheta,t}$ observing $[x_1, x_3]^\top$.}
		\label{fignmsetheta2states}
	\end{subfigure}
	%\hfill
	\begin{subfigure}{0.24\linewidth}
		\centering
		\input{EnKFvsAlg_obsxandz_MSExv3.tex}
		\caption{NMSE$_{x,t}$ observing $[x_1, x_3]^\top$.}
		\label{fignmsex2states}
	\end{subfigure}
	\caption{ Performance of \gls{ukf} (red), \gls{enkf} (blue) and \gls{ukf}-\glspl{ekf} (yellow) for two different setups, averaged over 50 independent simulation runs. Figures \ref{fignmsetheta3states} and \ref{fignmsetheta3states} show NMSE$_{\btheta,t}$ and NMSE$_{x,t}$ respectively, where the whole state vector is observed. In figures \ref{fignmsetheta2states} and \ref{fignmsex2states}, the error is plotted for a setup where only the first and third components of the state ($x_1$ and $x_3$) are observed. }
	\label{figcomparing}
\end{figure}

In the next experiment we compare the proposed nested Gaussian filters (Algorithm \ref{alNGFRecursive}) with two classical methods: the unscented Kalman filter (\gls{ukf}) \cite{Julier04} and the ensemble Kalman filter (\gls{enkf}) \cite{Evensen03}, both relying on the state-augmentation technique \cite{Liu01,Andrieu04} to incorporate the unknown parameters. To be specific, this approach implies that the system state $\bx_t$ is extended with the parameter vector to obtain the augmented state $\tilde{\bx}_t = \begin{bmatrix}
\bx_t \\
\btheta
\end{bmatrix}$. The \gls{ukf} and \gls{enkf} algorithms are used to track $\tilde{\bx}_t$ instead of $\bx_t$.

We have carried out two sets of computer simulations. In the first one we assume that the observation vectors are of the form $\by_t = k_o \bx_t + \br_t$, i.e., all the state variables are observed in Gaussian noise. The results are displayed in Fig. \ref{fignmsetheta3states}  and Fig. \ref{fignmsex3states}, which show the NMSE for the parameters $\btheta$ and the state $\bx_t$ over time, respectively, for the three competing algorithms. The nested scheme outperforms the augmented-state methods clearly in terms of parameter estimation (Fig. \ref{fignmsetheta3states}) and by a smaller margin in terms of state tracking (Fig. \ref{fignmsex3states}). When the observations are reduced to two state variables $\by_t = k_o \begin{bmatrix}
x_{1,t} \\
x_{3,t}
\end{bmatrix} + \br_t$, in Gaussian noise, the advantage of the nested scheme becomes larger, as shown in Figs. \ref{fignmsetheta2states} and \ref{fignmsex2states}.

%\textcolor{blue}{ Figure \ref{figcomparing} compares the average performance of Algorithm \ref{alNGFRecursive}, in yellow, with respect to ensemble Kalman filter (EnKF), in blue, and unscented Kalman filter (UKF), in red. In both algorithms we use state augmentation  doing prediction of an extended state vector that includes both state and parameters. Figures \ref{fignmsetheta3states} and \ref{fignmsex3states} show the NMSE$_{\btheta}$ and the NMSE$_x$ respectively for a simulation setup where the whole state vector is observed. For NMSE$_{\btheta}$, we obtain different results for each of the algorithms, being clearly better the performance of UKF-EKF, followed by EnKF. The algorithm proposed also obtained better results for the NMSE$_{x,t}$, however the difference with respect the other algorithms is not so remarkable. Although parameters estimation error of UKF and EnKF are clearly greater, the state estimation error remains low and quite close to the performance of UKF-EKF. On the other hand, when only a part of the state is observed, the state estimation error is not so similar among them, obtaining again a better performance with the proposed algorithm, followed by EnKF and UKF. The NMSE$_{\btheta,t}$ is slightly worse than in figure \ref{fignmsetheta3states} for each of the algorithms.}

\begin{figure}[htb]
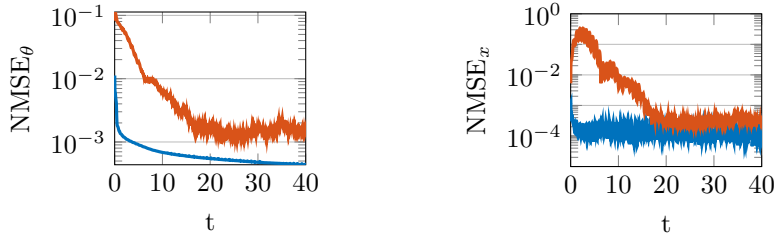

	\centering
	\begin{subfigure}{0.49\linewidth}
		\centering
		\input{SMCEKFvsUKFEKF_obsxandz_MSEthetav2.tex}
		\caption{Averaged NMSE$_{\btheta,t}$.}
		\label{figNHFnmsetheta}
	\end{subfigure}
	%\hfill
	\begin{subfigure}{0.49\linewidth}
		\centering
		\input{SMCEKFvsUKFEKF_obsxandz_MSExv2.tex}
		\caption{Averaged NMSE$_{x,t}$. }
		\label{figNHFnmsex}
	\end{subfigure}
	\caption{ Performance of a SMC-EKF (red) and a UKF-EKF (blue) averaged over 100 simulation runs. Figure \ref{figNHFnmsetheta} shows NMSE$_{\theta,t}$ and figure \ref{figNHFnmsex} shows NMSE$_{x,t}$.}  
	\label{figcomparingNHF}
\end{figure}

%\begin{table}
%	\begin{center}
%		\begin{tabular}{ |c|c|c|c| } 
%			\hline
%			& SMC-EKF & UKF-EKF \\
%			\hline
%			NMSE$_{x}$ & 3.3$\times 10^{-3}$ & 1.5$\times 10^{-3}$ \\ 
%			\hline
%			NMSE$_{\theta}$ & 1.5$\times 10^{-3}$ & 4.8$\times 10^{-4}$ \\ 
%			\hline
%			Running time in minutes & 14.8 & \textbf{4.5} \\ 
%			\hline
%		\end{tabular}
%		\caption{\label{tabNHFcomp} Average NMSE$_x$ and NMSE$_\theta$ of the last half simulation and average running time over 100 simulation runs, for SMC-EKF and UKF-EKF. }
%	\end{center}
%\end{table}

Next, we compare the performance of the UKF-EKF nested filter (Algorithm \ref{alNGFRecursive}) with one of the nested hybrid filters in \cite{Perez-Vieites18}. The latter method consists of a \gls{smc} filter with 120 particles for the first layer and a bank of \glspl{ekf} for the second layer. Figures \ref{figNHFnmsetheta} and \ref{figNHFnmsex} show the NMSE$_{\theta,t}$ and the NMSE$_{x,t}$ respectively, for both the SMC-EKF (red line) and the UKF-EKF (blue line) methods. 
Although the time of convergence of the SMC-EKF scheme can be reduced, the UKF-EKF algorithm converges clearly faster. Also, once it converges, the estimation error for both parameters and states is slightly lower for the UKF-EKF method. However, the greatest improvement is related to the computational cost. For this experiment %Table \ref{tabNHFcomp} not only summarizes the NMSE obtained for parameters and states, but also includes the average running time in minutes, showing that 
the UKF-EKF algorithm is three times faster (4.5 minutes run-time versus 14.8) than the SMC-EKF scheme. Therefore, it considerably reduces the computational cost while obtaining similar or slightly better results in estimation error.
%the error in parameters estimation is slightly greater for the SMC-EKF, the estimation error of the state variables is virtually the same for both schemes. It is also remarkable that the convergence of the algorithm is slower for the SMC-EKF. Table \ref{tabNHFcomp} summarizes this similar estimation errors but also shows that running time for schemes that uses Monte Carlo methods in the first layer is considerably greater (almost 15 minutes compared to the 5 minutes running time of UKF-EKF). 

\begin{figure*}[htb]
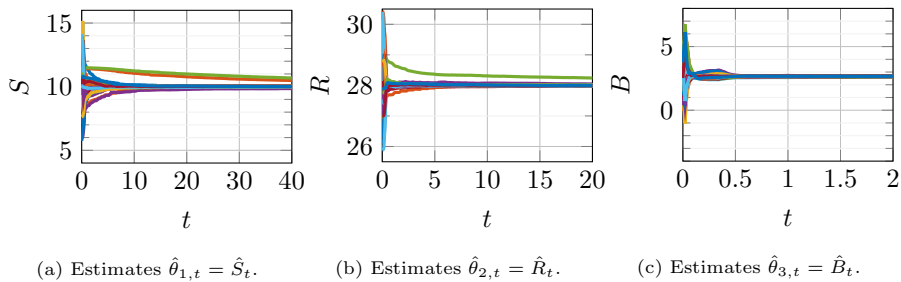

	\centering
	\begin{subfigure}{0.32\linewidth}
		\input{Strajv2.tex}
		\caption{Estimates $\hat{\theta}_{1,t} = \hat{S}_t$.}
		\label{figconvergenceS}
	\end{subfigure}
	%\hspace{0.1mm}
	\begin{subfigure}{0.32\linewidth}
		\input{Rtrajv2.tex}
		\caption{Estimates $\hat{\theta}_{2,t}= \hat{R}_t$.}
		\label{figconvergenceR}
	\end{subfigure}
	%\hspace{0.1mm}
	\begin{subfigure}{0.32\linewidth}
		\input{Btrajv2.tex}
		\caption{Estimates $\hat{\theta}_{3,t}= \hat{B}_t$.}
		\label{figconvergenceB}
	\end{subfigure}
	\caption{Sequences of posterior-mean estimates, $\hat{\btheta}_t$, over time obtained from $50$ independent simulation runs.}
	\label{figconvergence}
\end{figure*}

For the next computer experiment, Figure \ref{figconvergence}  shows the parameter estimates obtained by running 50 independent simulations of the proposed UKF-EKF nested filter. The three dimensions of $\hat{\btheta}_t$ are displayed over time (\ref{figconvergenceS}--\ref{figconvergenceB}) in order to illustrate how they converge as observations are collected. Although the length of the simulations is $T = 40$ continuous-time units, we have plotted just the intervals of time where the estimates converge. The interval varies from one plot to another because the time of convergence is not the same for all parameters (having shorter times for $B$ and longer times for $S$). In spite of that, this figure shows how all parameters converge to the true values for different initializations. 

%We have also assessed the accuracy of algorithm in terms of the mean square error (MSE) of the predictor of the state. We show the empirical MSE per dimension resulting directly from the simulations,
%\begin{equation}
%\text{NMSE}_t = \frac{\| \bx_t - \hat{\bx}_t \|^2}{\| \bx_t \|^2}. \label{eqMSE}
%\end{equation}

\begin{figure}[htb]
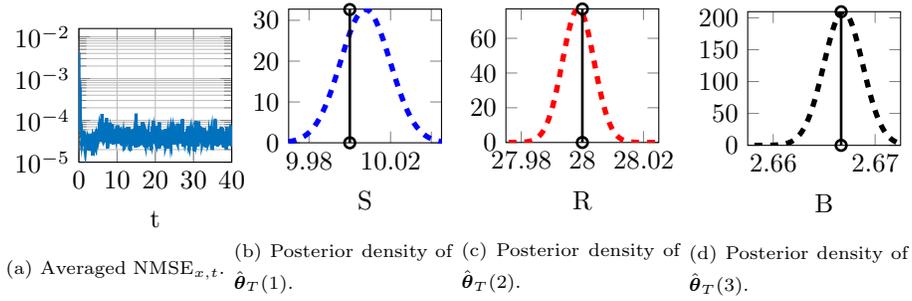

	\centering
	\begin{subfigure}{0.24\linewidth}
		\centering
		\input{meanMSExUKFEKFv2.tex}
		\caption{Averaged NMSE$_{x,t}$.}
		\label{figMSE}
	\end{subfigure}
	%\hfill
	\begin{subfigure}{0.24\linewidth}
		\centering
		\input{post_pdf_S_tfinal.tex}
		\caption{Posterior density of $\hat{\btheta}_T(1)$.}
		\label{figpdfS}
	\end{subfigure}
	\begin{subfigure}{0.24\linewidth}
		\centering
		\input{post_pdf_R_tfinal.tex}
		\caption{Posterior density of $\hat{\btheta}_T(2)$.}
		\label{figpdfR}
	\end{subfigure}
	%\hfill
	\begin{subfigure}{0.24\linewidth}
		\centering
\begin{tikzpicture}

\begin{axis}[%
width=0.8in,
height=0.7in,
at={(1.011in,0.642in)},
scale only axis,
xmin=2.6575,
xmax=2.6725,
xlabel={B},
ymin=0,
ymax=209.539500098643,
ylabel style= {yshift=-0.4cm},
xtick={2.66, 2.67},
axis background/.style={fill=white},
legend style={legend cell align=left,align=left,draw=white!15!black}
]
\addplot [color=black,dashed,line width=2.0pt,forget plot]
  table[row sep=crcr]{%
2.6566	9.9759507918192e-05\\
2.6567	0.000132247629714607\\
2.6568	0.000174833382257297\\
2.6569	0.000230496155250842\\
2.657	0.000303044129732682\\
2.6571	0.000397329608760264\\
2.6572	0.000519515904176497\\
2.6573	0.000677406920100195\\
2.6574	0.000880852668322752\\
2.6575	0.00114224636179634\\
2.6576	0.00147713148920748\\
2.6577	0.00190494040368829\\
2.6578	0.00244988948614591\\
2.6579	0.00314205988706325\\
2.658	0.00401869722159726\\
2.6581	0.00512576839235954\\
2.6582	0.00651981893077541\\
2.6583	0.00827017985323633\\
2.6584	0.0104615789743211\\
2.6585	0.0131972178340093\\
2.6586	0.0166023817786659\\
2.6587	0.0208286571540045\\
2.6588	0.0260588358530326\\
2.6589	0.0325125934031117\\
2.659	0.0404530321193844\\
2.6591	0.0501941852960226\\
2.6592	0.0621095816025096\\
2.6593	0.0766419704001577\\
2.6594	0.0943143081467112\\
2.6595	0.115742102922771\\
2.6596	0.141647207862915\\
2.6597	0.172873144345069\\
2.6598	0.210402021603215\\
2.6599	0.255373100396947\\
2.66	0.309103023935883\\
2.6601	0.373107708873653\\
2.6602	0.449125852398239\\
2.6603	0.539143967877682\\
2.6604	0.645422810922572\\
2.6605	0.770525000007527\\
2.6606	0.91734357106507\\
2.6607	1.08913113407353\\
2.6608	1.28952922221114\\
2.6609	1.52259734154758\\
2.661	1.79284114274349\\
2.6611	2.105239047432\\
2.6612	2.46526657281983\\
2.6613	2.8789175108904\\
2.6614	3.35272103612937\\
2.6615	3.89375374097271\\
2.6616	4.50964553450909\\
2.6617	5.20857829090895\\
2.6618	5.9992761033333\\
2.6619	6.89098599063527\\
2.662	7.89344792150272\\
2.6621	9.01685306761134\\
2.6622	10.2717892770866\\
2.6623	11.6691728748779\\
2.6624	13.2201660496605\\
2.6625	14.9360792790381\\
2.6626	16.8282584765987\\
2.6627	18.9079568150044\\
2.6628	21.1861914868236\\
2.6629	23.6735860060948\\
2.663	26.3801990240796\\
2.6631	29.3153410260987\\
2.6632	32.4873806851823\\
2.6633	35.903543063826\\
2.6634	39.5697022674217\\
2.6635	43.4901715506048\\
2.6636	47.6674942488279\\
2.6637	52.1022392395136\\
2.6638	56.792804918269\\
2.6639	61.7352358917396\\
2.664	66.9230567294394\\
2.6641	72.3471271710231\\
2.6642	77.9955231435036\\
2.6643	83.8534477973269\\
2.6644	89.9031765163117\\
2.6645	96.1240394918911\\
2.6646	102.492444977042\\
2.6647	108.981945753405\\
2.6648	115.563350663878\\
2.6649	122.204882292908\\
2.665	128.872381030894\\
2.6651	135.529554854305\\
2.6652	142.138273209081\\
2.6653	148.6589024234\\
2.6654	155.050679119849\\
2.6655	161.27211717098\\
2.6656	167.28144287162\\
2.6657	173.037052210713\\
2.6658	178.497983437902\\
2.6659	183.62439755726\\
2.666	188.378058961721\\
2.6661	192.72280816331\\
2.6662	196.625018485719\\
2.6663	200.054028676924\\
2.6664	202.982543671808\\
2.6665	205.386996185782\\
2.6666	207.247862442789\\
2.6667	208.549926122975\\
2.6668	209.282485539431\\
2.6669	209.439500098643\\
2.667	209.019673240984\\
2.6671	208.026470268398\\
2.6672	206.468070716431\\
2.6673	204.357256186195\\
2.6674	201.711235787733\\
2.6675	198.551412529193\\
2.6676	194.903095087214\\
2.6677	190.795160386727\\
2.6678	186.259673280037\\
2.6679	181.331470326316\\
2.668	176.047715217678\\
2.6681	170.447433768646\\
2.6682	164.571036575246\\
2.6683	158.459837458623\\
2.6684	152.155575640919\\
2.6685	145.699949267781\\
2.6686	139.134167405427\\
2.6687	132.498527017875\\
2.6688	125.832020692499\\
2.6689	119.171980052089\\
2.669	112.553758892738\\
2.6691	106.010459143967\\
2.6692	99.5727017858415\\
2.6693	93.2684439016415\\
2.6694	87.1228421159985\\
2.6695	81.1581617881546\\
2.6696	75.3937305167246\\
2.6697	69.8459337813683\\
2.6698	64.5282499100145\\
2.6699	59.4513210258183\\
2.67	54.623056203706\\
2.6701	50.0487627520401\\
2.6702	45.7313013308173\\
2.6703	41.6712605203554\\
2.6704	37.8671464573928\\
2.6705	34.3155832502436\\
2.6706	31.0115200611934\\
2.6707	27.9484409916319\\
2.6708	25.1185742112576\\
2.6709	22.5130971241867\\
2.671	20.1223347494425\\
2.6711	17.9359488990404\\
2.6712	15.9431161520351\\
2.6713	14.1326930365937\\
2.6714	12.4933672349486\\
2.6715	11.0137940098758\\
2.6716	9.68271740921134\\
2.6717	8.48907613131873\\
2.6718	7.42209422544695\\
2.6719	6.47135705385156\\
2.672	5.62687315658394\\
2.6721	4.87912283373732\\
2.6722	4.21909439572445\\
2.6723	3.63830913090282\\
2.6724	3.12883610423435\\
2.6725	2.68329793367846\\
2.6726	2.29486869615635\\
2.6727	1.95726509584426\\
2.6728	1.6647319880169\\
2.6729	1.41202329547762\\
2.673	1.19437928547319\\
2.6731	1.00750109635228\\
2.6732	0.84752331827005\\
2.6733	0.710985343864455\\
2.6734	0.594802115580509\\
2.6735	0.496234808351188\\
2.6736	0.412861901444867\\
2.6737	0.342551012900086\\
2.6738	0.283431795185771\\
2.6739	0.233870122313683\\
2.674	0.192443737070558\\
2.6741	0.15791947255814\\
2.6742	0.129232114864179\\
2.6743	0.105464933245466\\
2.6744	0.0858318703891552\\
2.6745	0.0696613577097534\\
2.6746	0.056381698734997\\
2.6747	0.0455079468824934\\
2.6748	0.0366301917541136\\
2.6749	0.0294031598937426\\
2.675	0.023537031192177\\
2.6751	0.0187893702316854\\
2.6752	0.0149580723287834\\
2.6753	0.0118752263855336\\
2.6754	0.00940180047120252\\
2.6755	0.00742306095050654\\
2.6756	0.00584464162478724\\
2.6757	0.00458918547903371\\
2.6758	0.00359348799693728\\
2.6759	0.00280607742710034\\
2.676	0.00218517370398945\\
2.6761	0.00169697383021445\\
2.6762	0.00131421732477136\\
2.6763	0.00101499077464831\\
2.6764	0.000781735556085746\\
2.6765	0.000600427396552048\\
2.6766	0.000459900623707333\\
2.6767	0.000351293699502631\\
};
\addplot [color=black,solid,line width=1.0pt,mark=o,mark options={solid},forget plot]
  table[row sep=crcr]{%
2.66666666666667	0\\
2.66666666666667	209.439500098643\\
};
\end{axis}
\end{tikzpicture}%
		\caption{Posterior density of $\hat{\btheta}_T(3)$.}
		\label{figpdfB}
	\end{subfigure}
	\caption{The mean NMSE$_{x,t}$ of $50$ simulation runs over time is plotted in \ref{figMSE}. Figures \ref{figpdfS}, \ref{figpdfR} and \ref{figpdfB} show the posterior density of parameters (dashed lines) at time $t = T$ and their true values (black vertical lines).}
	\label{figstatesMSE}
\end{figure}

\begin{figure}[htb]
	\centering
	\begin{subfigure}{0.49\linewidth}
		\centering
\definecolor{mycolor1}{rgb}{0.00000,0.44700,0.74100}%
\begin{tikzpicture}

\begin{axis}[%
width=1.1in,
height=0.8in,
at={(1.011in,0.642in)},
scale only axis,
xmode=log,
xmin=1,
xmax=10,
xminorticks=true,
xlabel={$\sigma_{y}^2$},
ymode=log,
ymin=7e-07,
ymax=2e-06,
yminorticks=true,
ymajorgrids,
yminorgrids,
xtick={1,2,4,10},
xticklabels={1,2,4,10},
ytick={7e-07,8e-07,9e-07,1e-06,2e-06},
yticklabels={,8$\times10^{-7}$,,$10^{-6}$,2$\times10^{-6}$},
ylabel={$\text{NMSE}_{\theta}$},
ylabel style= {yshift=0cm},
axis background/.style={fill=white},
legend style={legend cell align=left,align=left,draw=white!15!black}
]
\addplot [color=mycolor1,solid,mark=diamond,mark options={solid},line width=2.0pt,forget plot]
  table[row sep=crcr]{%
1	7.78223094650302e-07\\
2	7.78240236366647e-07\\
4	8.98355271219465e-07\\
10	1.1274209180172e-06\\
};
\end{axis}

%\begin{axis}[%
%width=7.778in,
%height=5.833in,
%at={(0in,0in)},
%scale only axis,
%xmin=0,
%xmax=1,
%ymin=0,
%ymax=1,
%hide axis,
%axis x line*=bottom,
%axis y line*=left,
%legend style={legend cell align=left,align=left,draw=white!15!black}
%]
%\end{axis}
\end{tikzpicture}%
		\caption{NMSE$_{\theta}$ for different $\sigma_{y}^2$.}
		\label{figchangingnoiseNMSEtheta}
	\end{subfigure}
	\hfill
	\begin{subfigure}{0.49\linewidth}
		\centering
\definecolor{mycolor1}{rgb}{0.00000,0.44700,0.74100}%
\begin{tikzpicture}

\begin{axis}[%
width=1.1in,
height=0.8in,
at={(1.011in,0.642in)},
scale only axis,
xmode=log,
xmin=1,
xmax=10,
xminorticks=true,
xlabel={$\sigma_{y}^{2}$},
ymode=log,
ylabel style= {yshift=0cm},
xticklabels={1,2,4,10},
ymin=2e-05,
ymax=2e-04,
xtick={1,2,4,10},
ytick={2e-05,3e-05,4e-05,5e-05,6e-05,7e-05,8e-05,9e-05,1e-04,2e-04},
yticklabels={2$\times10^{-5}$,,,,,,,,$10^{-4}$,2$\times10^{-4}$},
yminorticks=true,
ymajorgrids,
yminorgrids,
ylabel={$\text{NMSE}_{{x}}$},
axis background/.style={fill=white},
legend style={legend cell align=left,align=left,draw=white!15!black}
]
\addplot [color=mycolor1,solid,mark=diamond,mark options={solid},line width=2.0pt,forget plot]
  table[row sep=crcr]{%
1	3.84459040753736e-05\\
2	5.29860013270258e-05\\
4	7.27405451208744e-05\\
10	0.000120999436019415\\
};
\end{axis}
\end{tikzpicture}%
		\caption{NMSE$_x$ for different $\sigma_{y}^2$.}
		\label{figchangingnoiseNMSEx}
	\end{subfigure}
	\caption{NMSE$_{\theta}$ (\ref{figchangingnoiseNMSEtheta}) and NMSE$_x$ (\ref{figchangingnoiseNMSEx}) of UKF-EKF, averaged over 50 simulation runs, for different values of the noise variance $\sigma_{y}^2$.} 
	\label{figchangingnoise}
\end{figure}
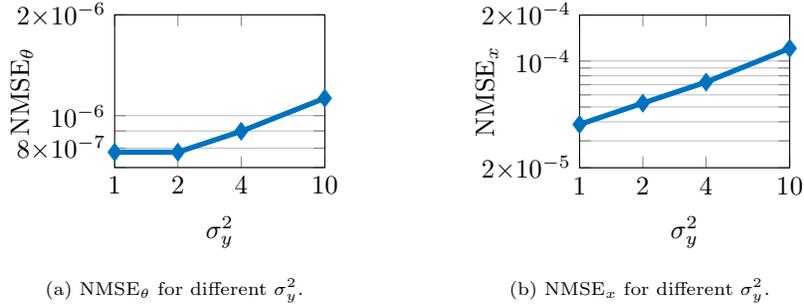

Figure \ref{figMSE}, on the other hand, illustrates the accuracy of state estimates, $\hat{\bx}_{t}$, by averaging the NMSE$_{x,t}$ obtained for the same set of 50 simulation runs as in Figure \ref{figconvergence}. The error NMSE$_{x,t}$ decreases with time as the parameter estimates get closer to their true values, being its value stabilized around $t= 5$. By that time, all parameter estimates in Figure \ref{figconvergence} have already converged (or at least got closer to their steady values) and, consequently, the state estimates become reliable.

In Figures \ref{figpdfS}, \ref{figpdfR} and \ref{figpdfB}, the estimated marginal \glspl{pdf} of each element in $\hat{\btheta}_t$ at time $t=40$ are plotted for a typical simulation run. These plot illustrate the uncertainty associated to each parameter. The means of these Gaussian \glspl{pdf} are close to the true parameters, in agreement with results seen in Figure \ref{figconvergence}. In addition, the variances are small, being all the probability distributions tightly packed around the ground truth.

Finally, Figure \ref{figchangingnoise} displays the average performance of the UKF-EKF nested filter for different observation noise variances, $\sigma_{y}^2$. Although all the previous experiments are done with $\sigma_{y}^2 = 1$, in figure \ref{figchangingnoiseNMSEtheta} we obtain similar results of NMSE$_{\theta}$ for $\sigma_{y}^2 = 2$ and slightly worse errors for $\sigma_{y}^2 = 4$ and $\sigma_{y}^2 = 10$. Although the errors increase for values of $\sigma_y^2$ greater than one, the general performance of the algorithm is still accurate for larger values of the variance in the observation noise.

%\begin{figure*}[htb]
%	\centering
%	\begin{subfigure}{0.32\linewidth}
%		\includegraphics[width=0.99\linewidth]{figures/postpdfSv11.eps}
%		\label{figpdfS}
%	\end{subfigure}
%	%\hspace{0.1mm}
%	\begin{subfigure}{0.32\linewidth}
%		\includegraphics[width=0.99\linewidth]{figures/postpdfRv11.eps}
%		\label{figpdfR}
%	\end{subfigure}
%	%\hspace{0.1mm}
%	\begin{subfigure}{0.32\linewidth}
%		\includegraphics[width=0.99\linewidth]{figures/postpdfBv11.eps}
%		\label{figpdfB}
%	\end{subfigure}
%	\caption{Posterior density of parameters (dashed lines) at time $t = 2 \times 10^4$ and their true values (in black).}
%	\label{figpdfs}
%\end{figure*}

\section{Conclusions}\label{sConclusions}
%We have introduce a nested filtering methodology that, based on nested hybrid filters (NHF) \cite{Sara18}, calibrates fixed parameters and estimates stochastic dynamical variables of a system using long sequences of observations collected over time. This scheme combines two layers of filtering, one inside the other, computing the joint posterior probability distribution of parameters and state. Specifically, we explore the use of a deterministic sampling technique such as unscented Kalman filter (UKF) \cite{Julier00} or cubature Kalman filter (CKF) \cite{Arasaratnam09} in the first layer for approximating the marginal posterior probability distribution of the parameters, while in the second layer, we calculate the approximate posterior probability distribution of the states using Gaussian filtering techniques. We avoid the use of Monte Carlo methods in any layer, making the algorithm well suited to high-dimensional state and parameter spaces. We have presented numerical results for a Lorenz 63 model, illustrating the average performance of parameter and state estimation.
We have introduced a generalization of the \gls{nhf} methodology of \cite{Perez-Vieites18} that, using long sequences of observations collected over time, estimates the static parameters and tracks the stochastic dynamical variables of a state space model. This scheme combines two layers of filters, one inside the other, in order to compute the joint posterior probability distribution of the parameters and the states. In this generalization of the methodology, we introduce the use of deterministic sampling techniques in the first layer of the algorithm (the cubature Kalman filter (CKF) or the unscented Kalman filter (UKF)), instead of Monte Carlo methods, describing in detail how the algorithms can work sequentially and recursively. We have presented numerical results for a stochastic Lorenz 63 model, using a scheme with an UKF for the parameters in the first layer, and \glspl{ekf} for the time-varying state variables in the second layer. We have introduced and assessed the values of a relative threshold that enables the algorithm to work recursively, and we have evaluated the performance of the algorithm in terms of the normalized mean square errors for the parameters and the dynamic state variables. We have also compared these results with other algorithms, such as the ensemble Kalman filter (EnKF) or the unscented Kalman filter (UKF), that implement state augmentation (i.e., an extended state that includes both parameters and state), and also with a \gls{nhf} with a \gls{smc} in the first layer with \glspl{ekf} in the second layer. The use of Gaussian filters in the two layers of the algorithm not only leads to a significant reduction in computational complexity compared to Monte Carlo-based implementations but also increases the accuracy compared to the state-augmented Gaussian filters.

%describes how to use Gaussian filters in all layers of the algorithm. This enables a significant reduction in computational complexity compared to Monte Carlo-based implementations, specially in problems with high-dimensional state and parameter spaces. Then, the new scheme allows the combination of virtually any type of Gaussian or particle filter in any of the two layers of the nested structure. Specifically, we show in detail how to obtain an NHF that employs a deterministic-sampling Gaussian approximation (the cubature Kalman filter (CKF) \cite{Arasaratnam09} or the unscented Kalman filter (UKF) \cite{Julier00}) in the parameter layer with an extended Kalman filter () in the second layer. We have presented numerical results for a Lorenz 63 model.
\section{Acknowledgment}
This research was partially supported by \textit{Agencia Estatal de Investigaci\'on of Spain} (RTI2018-099655-BI00 CLARA), the regional government of Madrid (project no. Y2018/TCS-4705 PRACTICO) and the Office of Naval Research (award no. N0014-19-1-2226).

\appendix

\section{Proof of Proposition \ref{prop1}} \label{approof}

We proceed by induction in the time index $t$. For $t=0$ we have $\pi_0(\bx_0|\btheta)=p(\bx_0)$ independently of $\btheta$, hence for any pair $(\btheta,\btheta')\in\Theta\times\Theta$ we obtain
\begin{equation}
\int \big|\pi_0(\bx_0|\btheta)-\pi_0(\bx_0|\btheta')\big| d\bx_0 = \int \big|p(\bx_0)-p(\bx_0')\big| = L_0 \|\btheta-\btheta'\| \nonumber
\end{equation}
for $L_0=0$.

For the induction step, assume that
\begin{equation}
\int \big|\pi_{t-1}(\bx_{t-1}|\btheta) - \pi_{t-1}(\bx_{t-1}\big|\btheta')| d\bx_{t-1} < L_{t-1} \|\btheta - \btheta'\| \label{eqap0}
\end{equation}
for some $L_{t-1} < \infty$. Straightforward calculations yield
\begin{align}
\int& \big|\xi_t(\bx_t|\btheta) - \xi_t(\bx_t|\btheta')\big| d\bx_t = \nonumber \\
&=\int \bigg\vert \int p(\bx_t|\bx_{t-1},\btheta) \pi_{t-1}(\bx_{t-1}|\btheta)d\bx_{t-1} - \int p(\bx_t | \bx_{t-1},\btheta')\pi_{t-1}(\bx_{t-1}|\btheta')d\bx_{t-1} \nonumber \\
\pm& \int p(\bx_t|\bx_{t-1}, \btheta)\pi_{t-1}(\bx_{t-1}|\btheta')d\bx_{t-1} \bigg\vert d\bx_t \nonumber \\
%&= \int \bigg\vert \int \Big( p(\bx_t|\bx_{t-1},\btheta) \big[ \pi_{t-1}(\bx_{t-1}|\btheta) - \pi_{t-1}(\bx_{t-1}|\btheta') \big] \nonumber \\
%+& \big[ p(\bx_t|\bx_{t-1},\btheta) - p(\bx_t|\bx_{t-1},\btheta') \big] \pi_{t-1}(\bx_{t-1}|\btheta') \Big) d\bx_{t-1} \bigg\vert d\bx_t \nonumber \\
\le& \int \int p(\bx_t|\bx_{t-1},\btheta) \big\vert \pi_{t-1}(\bx_{t-1}|\btheta) - \pi_{t-1}(\bx_{t-1}|\btheta') \big\vert d\bx_{t-1} d\bx_t \nonumber \\
+& \int \int \big\vert p(\bx_t |\bx_{t-1},\btheta) - p(\bx_t|\bx_{t-1},\btheta') \big\vert \pi_{t-1}(\bx_{t-1}|\btheta') d\bx_{t-1} d\bx_t \nonumber %\label{eqintxi1} 
\end{align}
and reordering the integrals we obtain
\begin{align}
\int& \big|\xi_t(\bx_t|\btheta) - \xi_t(\bx_t|\btheta')\big| d\bx_t \le \nonumber\\
&\le \int \bigg[\int p(\bx_t|\bx_{t-1},\btheta) d\bx_t \bigg] \big\vert \pi_{t-1}(\bx_{t-1}|\btheta) - \pi_{t-1}(\bx_{t-1}|\btheta') \big\vert d\bx_{t-1}  \nonumber \\
+& \int \bigg[ \int \big\vert p(\bx_t |\bx_{t-1},\btheta) - p(\bx_t|\bx_{t-1},\btheta') \big\vert d\bx_t \bigg] \pi_{t-1}(\bx_{t-1}|\btheta') d\bx_{t-1}  \label{eqap1}
\end{align}
However, $\int p(\bx_t|\bx_{t-1},\btheta) d\bx_t=1$ for any $\bx_{t-1}$ and any $\btheta$, while Assumption \ref{as2} yields $\int \vert p(\bx_t|\bx_{t-1},\btheta) - p(\bx_t|\bx_{t-1},\btheta') \vert d\bx_t \le L \|\btheta-\btheta'\|$. Therefore, \eqref{eqap1} becomes
\begin{align}
\int \big|\xi_t(\bx_t|\btheta) - \xi_t(\bx_t|\btheta')\big| d\bx_t &\le \int \big\vert \pi_{t-1}(\bx_{t-1}|\btheta)- \pi_{t-1}(\bx_{t-1}|\btheta') \big\vert d\bx_{t-1} \nonumber\\
+& L \|\btheta-\btheta'\| \int \pi_{t-1}(\bx_{t-1}|\btheta')d\bx_{t-1} \nonumber \\
 & \le (L_{t-1} + L) \|\btheta-\btheta'\| \label{eqap1_5}
\end{align}
where the second inequality follows from the induction hypothesis \eqref{eqap0}.

As for the difference between $\pi_t(\cdot|\btheta)$ and $\pi_t(\cdot|\btheta')$, the Bayes theorem readily yields
\begin{align}
\int \big\vert \pi_t(\bx_t|\btheta) - \pi_t(\bx_t|\btheta') \big\vert d\bx_t = \int \bigg\vert \frac{p(\by_t|\bx_t,\btheta)\xi_t(\bx_t|\btheta)}{\eta_t(\by_t|\btheta)} - \frac{p(\by_t|\bx_t,\btheta')\xi_t(\bx_t|\btheta')}{\eta_t(\by_t|\btheta')} \bigg\vert d\bx_t \label{eqap2}
\end{align}
and the absolute difference in the integrand of \eqref{eqap2} can be rewritten as
\begin{align}
\bigg\vert& \frac{p(\by_t|\bx_t,\btheta)\xi_t(\bx_t|\btheta)}{\eta_t(\by_t|\btheta)} - \frac{p(\by_t|\bx_t,\btheta')\xi_t(\bx_t|\btheta')}{\eta_t(\by_t|\btheta')} \bigg\vert \nonumber \\
&= \bigg\vert \frac{p(\by_t|\bx_t,\btheta)\xi_t(\bx_t|\btheta)}{\eta_t(\by_t|\btheta)} \pm \frac{p(\by_t|\bx_t, \btheta')\xi_t(\bx_t|\btheta')}{\eta_t(\by_t|\btheta)} - \frac{p(\by_t|\bx_t,\btheta')\xi_t(\bx_t|\btheta')}{\eta_t(\by_t|\btheta')} \bigg\vert \nonumber \\
&= \bigg\vert \frac{p(\by_t|\bx_t,\btheta)\xi_t(\bx_t|\btheta)-p(\by_t|\bx_t,\btheta')\xi_t(\bx_t|\btheta')}{\eta_t(\by_t|\btheta)} + \pi_t(\bx_t|\btheta') \frac{\eta_t(\by_t|\btheta')- \eta_t(\by_t|\btheta)}{\eta_t(\by_t|\btheta)} \bigg\vert \label{eqap3}
\end{align}
where we have used the relationship $\pi_t(\bx_t|\theta')=\frac{p(\by_t|\bx_t,\btheta')\xi_t(\bx_t|\btheta)}{\eta_t(\by_t|\btheta')}$ to obtain the second identity. Now, if we substitute \eqref{eqap3} into \eqref{eqap2} and then realize that
\begin{equation}
\big\vert \eta_t(\by_t|\btheta) - \eta_t(\by_t|\btheta') \big\vert \le \int \big| p(\by_t|\bx_t,\btheta) \xi_t(\bx_t|\btheta) - p(\by_t|\bx_t,\btheta')\xi_t(\bx_t|\btheta') \big| d\bx_t \nonumber
\end{equation}
and $\int \pi_t(\bx_t|\btheta') d\bx_t=1$, we obtain the upper bounds
\begin{align}
\int& \big| \pi_t(\bx_t|\btheta)- \pi_t(\bx_t|\btheta') \big| d\bx_t \nonumber \\
&\le \frac{2}{\eta_t(\by_t|\btheta)} \int \bigg| p(\by_t|\bx_t,\btheta)\xi_t(\bx_t|\btheta) - p(\by_t|\bx_t,\btheta')\xi_t(\bx_t|\btheta') \bigg| d\bx_t \label{eqap4} \\
&\le \frac{2}{\eta_t(\by_t|\btheta)} \bigg[ \int p(\by_t|\bx_t,\btheta) \big| \xi_t(\bx_t|\btheta) - \xi_t(\bx_t|\btheta') \big| d\bx_t  \nonumber \\
+& \int \big| p(\by_t|\bx_t,\btheta) - p(\by_t|\bx_t,\btheta') \big| \xi_t(\bx_t|\btheta') d\bx_t \bigg] \label{eqap5}
\end{align}
where \eqref{eqap5} is obtained by applying a triangular inequality in \eqref{eqap4}.

The first integral in \eqref{eqap5} can be bounded using Assumption \ref{as4} and inequality \eqref{eqap1_5}, which together yield, 
\begin{align}
	\int \frac{p(\by_t|\bx_t,\btheta)}{\eta_t(\by_t|\btheta)} \big| \xi_t(\bx_t|\btheta) - \xi_t(\bx_t|\btheta') \big| d\bx_t &\le M_t \int \bigg| \xi_t(\bx_t|\btheta) - \xi_t(\bx_t|\btheta) \bigg| d\bx_t \nonumber \\
	&\le M_t (L_{t-1} + L) \| \btheta - \btheta'\|, \label{eqap6}
\end{align}
while the second integral can be bounded using Assumption \ref{as3}, which leads to
\begin{align}
\frac{2}{\eta_t(\by_t|\btheta)} \int \big| p(\by_t|\bx_t,\btheta) - p(\by_t|\bx_t,\btheta') \big| \xi_t(\bx_t|\btheta') d \bx_t &\le 2 G_t \|\btheta-\btheta'\| \int \xi_t (\bx_t|\btheta') d\bx_t \nonumber\\
&= 2 G_t \|\btheta-\btheta'\|. \label{eqap7}
\end{align}
Plugging \eqref{eqap6} and \eqref{eqap7} into \eqref{eqap5} yields
\begin{align}
\int \big| \pi_t(\bx_t|\btheta) - \pi_t(\bx_t|\btheta') \big| &\le M_t(L_{t-1}+L) \|\btheta-\btheta'\| + 2 G_t \|\btheta-\btheta'\| \nonumber \\
&\le L_t \|\btheta - \btheta'\| \nonumber
\end{align}
with $L_t = M_t (L_{t-1}+L) + 2 G_t < \infty$. $\QED$

\section*{References}

\bibliographystyle{elsarticle-num}
% \bibliography{bibliography}
\bibliography{bibliografia}

\end{document}